\documentclass[submitting]{nst}
\usepackage{subfigure,dcolumn}
\usepackage[T2A,T1]{fontenc}
\usepackage[russian,english]{babel}
\usepackage{placeins}
\usepackage{listings}
\usepackage{graphicx}
\usepackage{subfigure}
\usepackage{float}
\usepackage{stfloats}
\usepackage{geometry}
\usepackage{afterpage}
\usepackage{titlesec}
\usepackage{makecell}
\titleformat{\section}{\normalfont\Large\bfseries\raggedright}{\thesection}{1em}{}   
\titleformat{\subsection}{\normalfont\bfseries\raggedright}{\thesubsection}{1em}{}    
\titleformat{\subsubsection}{\normalfont\bfseries\raggedright}{\thesubsubsection}{1em}{}     
\renewcommand\thesection{\arabic{section}}
\renewcommand\thesubsection{\arabic{section}.\arabic{subsection}}
\renewcommand\thesubsubsection{\arabic{section}.\arabic{subsection}.\arabic{subsubsection}}

\begin{document}

\title{Radiation tolerance test and damage of single-crystal CVD Diamond sensor under high fluence particles}\thanks{Supported by the International Science \& Technology Cooperation Program of China (No. 2015DFG02100), the Ministry of Science and Technology of the People's Republic of China.}

\author{Jialiang Zhang}
\email[Contact author, Jialiang Zhang: ]{jlzhang@smail.nju.edu.cn}
\affiliation{National Laboratory of Solid State Microstructures, Nanjing University, Nanjing 210093, China}
\affiliation{School of Physics, Nanjing University, Nanjing 210093, China}
\author{Shuo Li}
\affiliation{National Laboratory of Solid State Microstructures, Nanjing University, Nanjing 210093, China}
\affiliation{School of Physics, Nanjing University, Nanjing 210093, China}
\author{Yilun Wang}
\affiliation{National Laboratory of Solid State Microstructures, Nanjing University, Nanjing 210093, China}
\affiliation{School of Physics, Nanjing University, Nanjing 210093, China}
\author{Shuxian Liu}
\affiliation{National Laboratory of Solid State Microstructures, Nanjing University, Nanjing 210093, China}
\affiliation{School of Physics, Nanjing University, Nanjing 210093, China}
\author{Guojun Yu}
\affiliation{National Laboratory of Solid State Microstructures, Nanjing University, Nanjing 210093, China}
\affiliation{School of Physics, Nanjing University, Nanjing 210093, China}
\author{Zifeng Xu}
\affiliation{National Laboratory of Solid State Microstructures, Nanjing University, Nanjing 210093, China}
\affiliation{School of Physics, Nanjing University, Nanjing 210093, China}
\author{Lifu Hei}
\affiliation{School of Materials Science and Engineering, University of Science and Technology Beijing, Beijing 100083, China}
\author{Fanxiu Lv}
\affiliation{School of Materials Science and Engineering, University of Science and Technology Beijing, Beijing 100083, China}
\author{Lei Zhang}
\affiliation{National Laboratory of Solid State Microstructures, Nanjing University, Nanjing 210093, China}
\affiliation{School of Physics, Nanjing University, Nanjing 210093, China}
\author{Ming Qi}
\email[Corresponding author, Ming Qi: ]{qming@nju.edu.cn}
\affiliation{National Laboratory of Solid State Microstructures, Nanjing University, Nanjing 210093, China}
\affiliation{School of Physics, Nanjing University, Nanjing 210093, China}

\begin{abstract}
Single-crystal chemical vapor deposition (CVD) diamond is a promising material for radiation detectors operating in extreme environments, owing to its outstanding radiation hardness. As nuclear and high-energy physics applications demand particle detectors that withstand higher radiation fluences, understanding the damage thresholds and degradation mechanisms of diamond-based detectors is essential for their practical operation.
In this study, Synthetic single-crystal CVD diamond sensors were exposed to fast neutron irradiation at fluences up to \SI{3.3e17}{n/cm^2}, one of the highest test doses for evaluating radiation tolerance in diamond detectors. 
Modules exhibited stable signal output, retaining approximately 5\% of their initial response after irradiation, confirming potential for application in future high-dose radiation environments.
Fast neutron induced damage in the diamond crystals was characterized using photoluminescence and scanning electron microscopy. The dominant defects were identified as point defects including $\langle 100\rangle$ self interstitials, vacancies, and lattice disorder. In addition, macroscopic defects on the crystal surface, including nanocavities and cracks, were observed with areal densities approaching \SI{e7}{cm^{-2}}.
The impact of \SI{100}{MeV} proton irradiation on diamond detector response was quantified by extracting a damage constant of $k^{\SI{100}{MeV}}_{\mathrm{proton}} = (1.452 \pm 0.006) \times 10^{-18}~\mathrm{cm^2/(p\cdot\mu m)}$ from a linear carrier drift degradation model.
Moreover, the mean free path of carriers was found to exhibit saturation behavior beyond a fluence of \SI{4e16}{p/cm^2} under \SI{100}{MeV} proton irradiation.
Monte Carlo together with molecular dynamics simulations were performed to assess irradiation induced defect production and its influence on carrier transport. The results indicate that saturation arises when local frenkel defect densities exceed \SI{e18}{/cm^{-3}}, at which defect interactions and clustering begin to dominate during irradiation.
By considering saturation effects and defect-interaction corrections, we develop an enhanced carrier-drift degradation model that accurately captures detector response under high-dose irradiation. Furthermore, the simulation framework was applied to evaluate damage induced by protons and pions on diamond at various energies, yielding results that show better agreement with experimental data than conventional NIEL based estimates. 
\end{abstract}


\maketitle

\tableofcontents

\section{Introduction}
Over the past decades, particle physics has entered a transformative era, marked by unprecedented precision and discovery. The 2012 observation of the Higgs boson \cite{bib:1, bib:2} signaled a milestone, pushing experimental physics toward probing the Standard Model at increasingly finer scales while exploring potential new physics beyond it. These pursuits have driven accelerator and collider upgrades along both the luminosity and energy frontiers as well as the foundation of conceptual design for next-generation experiments\cite{bib:3}, setting the stage for demanding operational conditions where detector radiation tolerance becomes critical\cite{bib:4}. In parallel, applications in nuclear fusion systems are approaching environments of comparable or even greater radiation severity\cite{bib:5}, further amplifying the need for robust detection technologies.

Among candidate materials, synthetic single-crystal diamond has emerged as a leading contender for next-generation radiation detectors. It combines exceptional material properties, including high charge carrier mobilities large than \SI{3000}{cm^2/(V\cdot s)}\cite{bib:6}, a wide bandgap of 5.45 eV that enables excellent signal-to-noise performance, and superior thermal conductivity reaching \SI{2000}{W/(m\cdot K)}\cite{bib:7}. Most critically, its radiation hardness surpasses silicon by factors of three for low-energy incident particles and by more than a factor of ten at high energies\cite{bib:8,bib:9}, which is attributed to its high displacement energy of up to 43.5 eV\cite{bib:10}. The advent of high-quality chemical vapor deposition (CVD) techniques has made large-area, defect-minimized single-crystal diamond (scCVD) substrates increasingly accessible\cite{bib:11,bib:12}, 
making their a promising choice for applications in detectors subjected to high radiation levels.
scCVD diamond detectors have been successfully employed in previous and ongoing experiments and facilities for beam monitoring and particle tracking, including the LHC, SuperKEKB, and the EAST tokamak system\cite{bib:13,bib:14,bib:15,bib:16}.

Despite these advances, the upper radiation dose limits that scCVD diamond detectors can tolerate without critical signal loss remain an open question. Prior studies, including work from RD42\cite{bib:17,bib:18}, have demonstrated the viability of diamond detectors under proton and neutron fluences approaching \SIrange{e15}{e16}{\per\cm\squared}. However, future environments are expected to push beyond these thresholds\cite{bib:19}. Moreover, irradiation sources often include mixed particle fields with broad energy spectra, making it essential to understand not only how damage accumulates but also how it impacts charge transport at the microscopic level and consequently affects detector performance.
Detector performance degradation is commonly understood as a two-step process: irradiation introduces lattice defects, and these defects in turn act as traps or recombination centers, reducing carrier lifetimes and thus degrades charge collection efficiency and detector performance\cite{bib:20}.

Although diamond is renowned for its exceptional radiation hardness, a detailed understanding of how irradiation-induced defects influence charge transport is still lacking. To address this gap, pioneering studies have carried out systematic irradiation experiments using protons and pions at multiple energies. These efforts produced quantitative data on signal degradation, which in turn enabled the development of simplified damage models\cite{bib:18,bib:21}. Such models have been applied to comparatively assess the effects of different particles and energy levels on diamond detector degradation\cite{bib:22,bib:23,bib:24}.
Beyond these experimental insights, predictive understanding requires complementary modeling approaches. Two main strategies have been developed. The first relies on the concept of non-ionizing energy loss (NIEL)\cite{bib:25}, where the energy deposited into the lattice leads to atomic displacements and crystal damage. The NIEL cross section provides a convenient and widely used metric for estimating radiation damage. The second strategy adopts a material-centric perspective, explicitly evaluating the number and types of defects generated in the lattice\cite{bib:26,bib:27}. This defect-informed methodology offers a more physically grounded framework for predicting detector degradation.

In this study, we present a comprehensive investigation into the radiation tolerance and damage mechanisms  of scCVD diamond sensors exposed to high fluence fast neutron and protons. 
Four scCVD diamond sensors were fabricated and irradiated at nuclear facilities for a fast neutron radiation experiment. The cumulative neutron fluence achieved \SI{3.3e17}{n\per\cm\squared}, one of the highest doses reported for single-crystal diamond to date. The sensors exhibited a sustained signal response, underscoring the potential of scCVD diamond as a viable candidate for mitigating the limited operational lifetime of silicon-based detectors, especially in the innermost layers of next-generation harsh radiation experiments. 
To elucidate the microscopic origins of radiation-induced performance degradation, the neutron irradiated diamonds were systematically characterized using photoluminescence spectroscopy and scanning electron microscopy. Damage was observed to generate atomic-scale point defects, crystalline lattice disorder, and macroscopic defects including voids and microcracks.

To assess the correlation between defect generation and transport degradation, we analyzed signal response data from scCVD diamond detectors previously irradiated with 100 MeV protons. Using a simplified carrier drift degradation model, we extracted a damage constant and normalized it to radiation damage of 24 GeV protons via established scaling relations\cite{bib:28}, allowing direct comparison with results from other irradiation studies. This analysis contributes quantitative insight into an energy regime that has remained relatively unexplored. Notably, we observed that the carrier mean free path exhibited saturation behavior at high fluence levels, suggesting a shift in the dominant damage mechanisms.
To interpret this effect, a combined simulation framework including Monte Carlo simulations and molecular dynamics modeling incorporating adiabatic recombination (arc-DPA)\cite{bib:27} were employed. The results indicate that when local defect densities exceed, interactions among defects begin to dominate over isolated point defect formation, driving a transition toward saturation in performance loss. Based on this, we refined the traditional carrier drift degradation model to account for saturation effects at extreme doses.
Finally, the combined simulation framework was applied to assess radiation damage from protons and pions across a range of energies. The predictions show closer agreement with experimental data than conventional NIEL-based estimates, highlighting the importance of defect-level modeling for accurate performance forecasting. Together, these results expand our understanding of diamond detector behavior under extreme radiation and provide actionable insights for their deployment in high-luminosity colliders, nuclear science such as fusion reactors, and space-based instruments.

\section{Experiment and Methods}\label{Experiment and Methods}
\subsection{Fabrication of DUT modules}
\begin{figure}[!t]
\centering
\subfigure[]{
  \includegraphics[width=0.9\hsize]{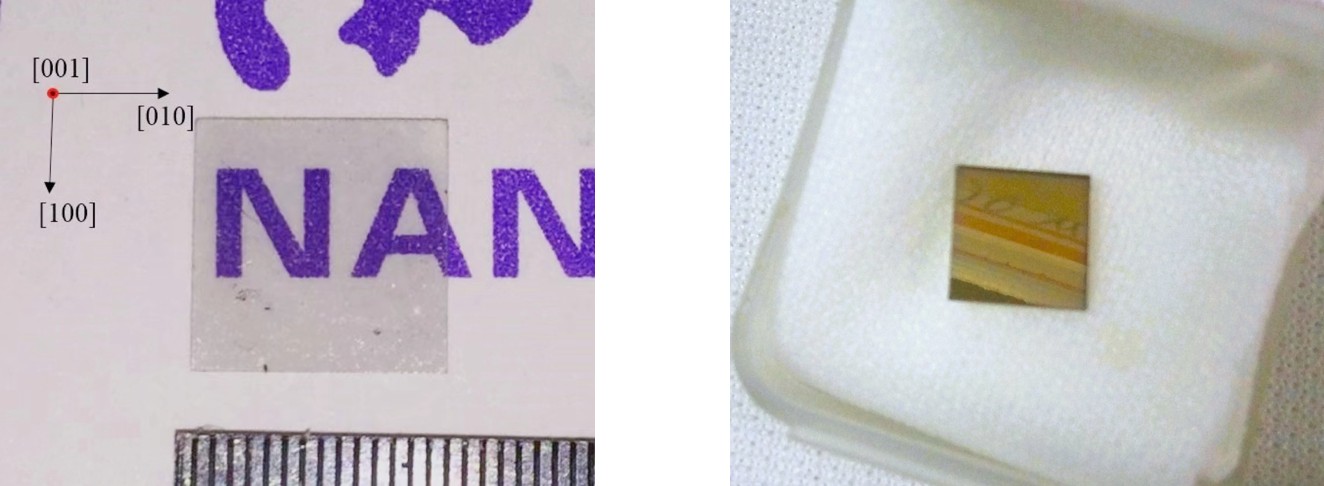}
}
\subfigure[]{
  \includegraphics[width=\hsize]{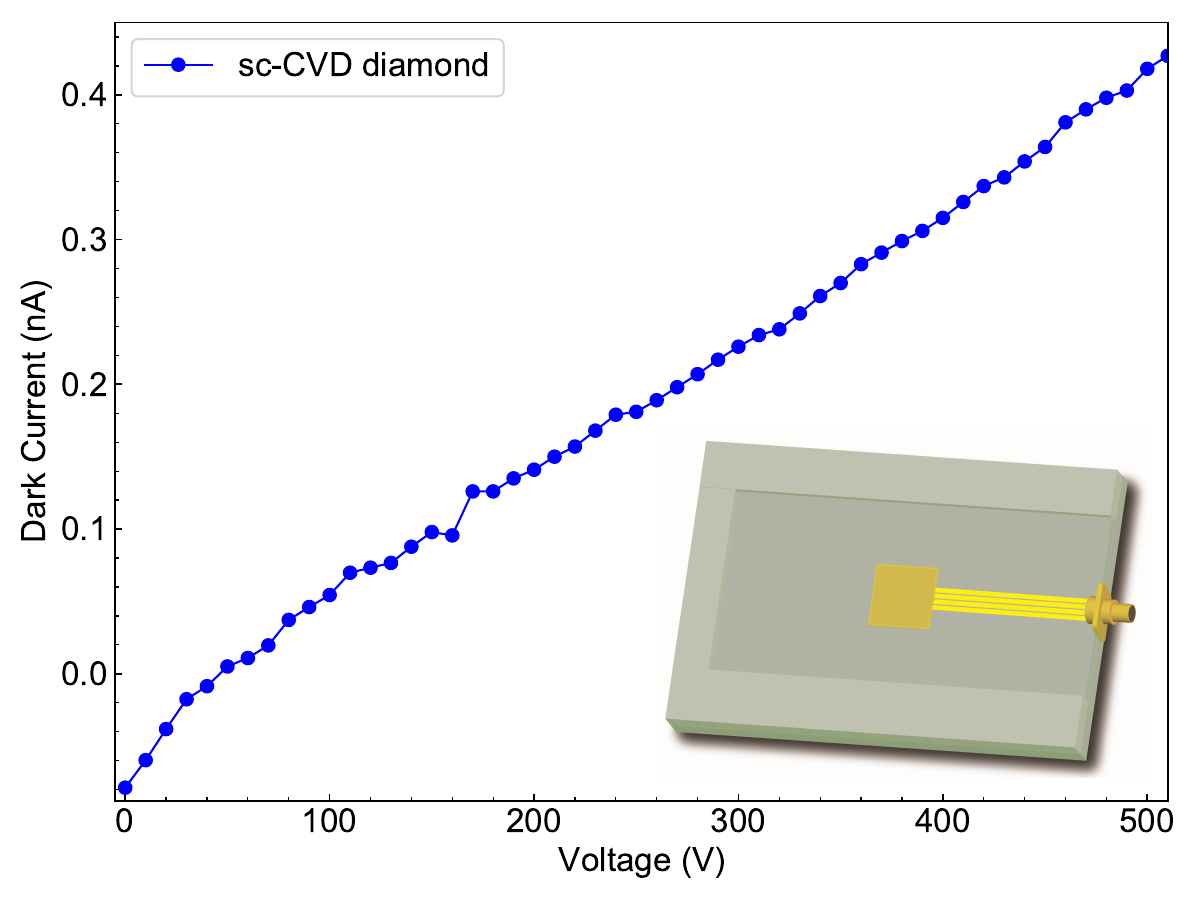}
}
\caption{(a) Optical micrograph of a synthetic scCVD diamond plate and the metallized sensor. (b) I-V characteristics and the assembled detector module mounted on a Rogers PCB with wire-bonded connections.}
\label{fig1}
\end{figure}

The single crystal diamonds material were synthetic using a commercial \SI{30}{\kW} DC arc plasma jet chemical vapor deposition (CVD) system operated in gas recycling mode. On the substrates of commercial high pressure high temperature (HPHT) type-Ib (100) single-crystal diamond, a number of high quality large-sized single-crystal diamond plates were fabricated using the CVD homoepitaxial growth technique. In preparation for the fast neutron irradiation experiment, self-supporting synthetic scCVD diamond plates were separated from the substrates by laser cutting. Then, mechanical polishing and boiling with a combination of acids were also used in order to get rid of any potential contaminations and damaged layers on the surfaces of scCVD diamond. Following the cutting, polishing, and cleaning procedures, scCVD diamond plates have the final size with a surface area of $7.0 \times 7.0$ \SI{}{\mm^2} and thickness around \SI{300}{\um}, as depicted in Fig.~\ref{fig1}(a). Raman spectroscopy was carried out on the surfaces of the plates to examine the purity and perfection of the single-crystal diamond, the results\cite{bib:29} of strong first-order peak at \SI{1332}{\cm^{-1}} with a narrow full wave at half maximum (FWHM) of \SI{2}{\cm^{-1}} demonstrating the obtain of high quality synthetic scCVD diamond plates prepared for radiation detection sensors. Planar Ti–W–Au electrodes were deposited on both surfaces of the scCVD diamond plates by magnetron sputtering, forming efficient metal–insulator–metal (MIM) detection sensors used as the Device Under Test (DUT) sensors, as illustrated in Fig.~\ref{fig1}(a). The electrical properties of these sensors were assessed through I-V curve measurements, as depicted in Fig.~\ref{fig1}(b). The curve reveals a minimal dark current of approximately 0.4 nA under a 500 V bias voltage with good linearity, indicating robust ohmic contact between the electrodes and the diamond. Subsequently,
DUT sensors were incorporated into Rogers ceramic base high-frequency PCBs as modules, featuring planar electrodes connected to the readout electronics through gold wire bonding, as show in Fig.~\ref{fig1}(b). Long Kapton-insulated coaxial cables were utilized for electronic communication, facilitating a connection to the remote data acquisition (DAQ) system.

\subsection{Radiation tolerance experiment}
\begin{figure}[!t]
\centering
\subfigure[]{
  \includegraphics[width=0.9\hsize]{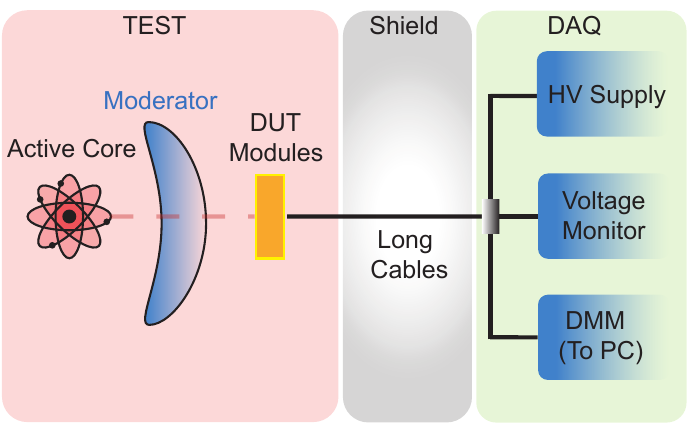}
}
\subfigure[]{
  \includegraphics[width=\hsize]{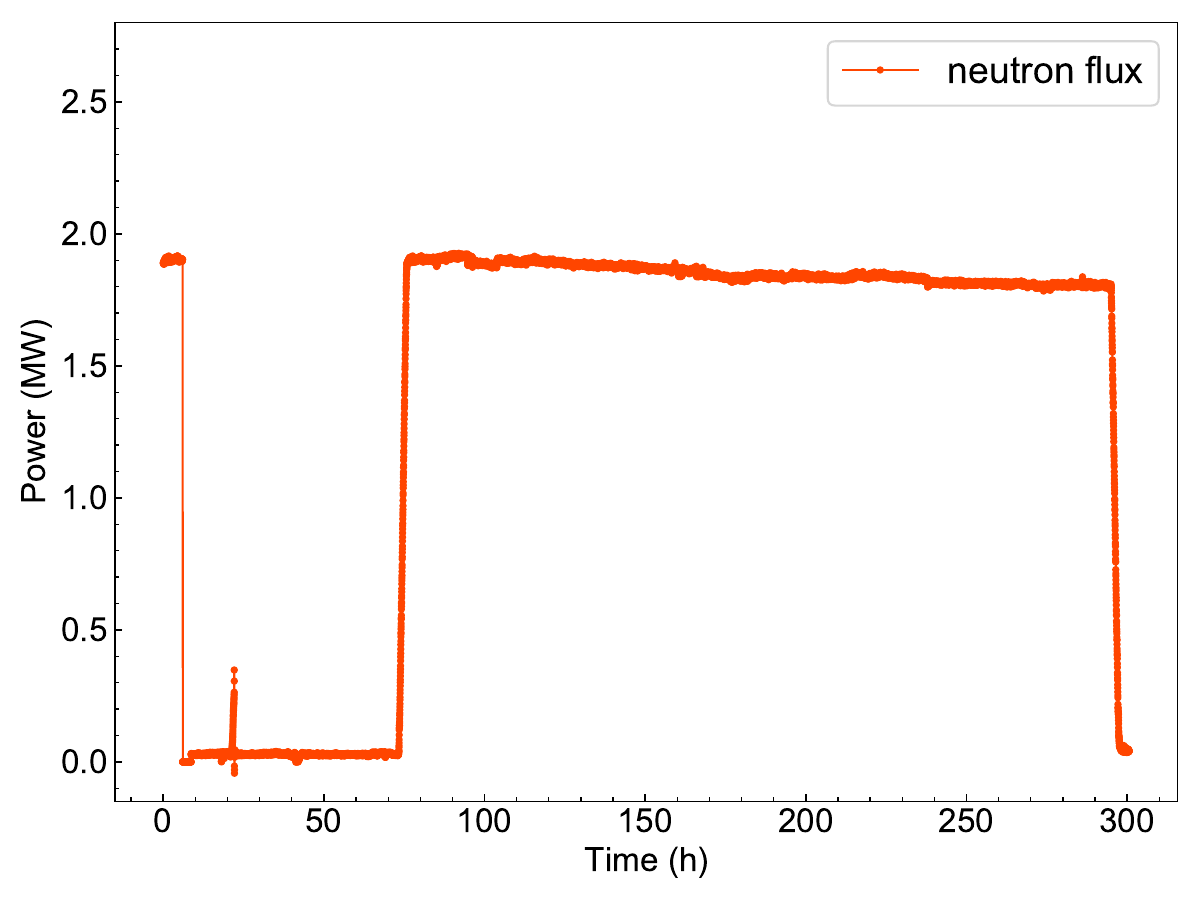}
}
\caption{Fast neutron irradiation experiment. (a) Schematic of the diamond DUT exposure to neutron beam. (b) Reactor power profile of the IBR-2M facility during testing.}
\label{fig2}
\end{figure}

The fast neutron irradiation was undertaken at the IBR-2M reactor in the Laboratory of Neutron Physics, Joint Institute for Nuclear Research, Dubna, Russia. The experiment took place in the beamline specifically designed for neutron irradiation experiments\cite{bib:30}. The DUT modules were placed inside a container, which was positioned at one end of an extended conduit and located 30 centimeters away from the reactor moderator, to get high fluence of fast neutrons. Connected by long cables passing through a nuclear radiation shielded area, DUT modules were linked to the DAQ system. The DAQ system consists of a high voltage supply module, a voltage monitoring module, and a digital multimeter. The high-voltage supply module is responsible for providing and monitoring bias voltage up to 260 V on the DUT sensors during the experiment, supplied by the Keithley 6487 module. The multi-channel multi-meter Keithley 2700, equipped with the Keithley 7703 relay card, is used to read out and record DC current ionization detection signals triggered by fast neutrons, and transmit them to the computer. The schematic diagram of the entire process is shown in Fig.~\ref{fig2}(a). One HV line, not connected to the detector, was also read out to measure dark current and noise. The entire irradiation experiment accumulated for approximately 280 hours. During this period, the reactor’s operating power was maintained at around \SI{2}{\MW}, as shown in Fig.~\ref{fig2}(b). The accumulated fast neutron irradiation fluence at the location of DUT modules reached \SI{3.3e17}{n/cm^2}, with a flux of \SI{3.27e11}{n\per\cm\squared\per\s}.

\section{Results and discussion}
\subsection{Radiation tolerance of scCVD diamond detector}
\begin{figure}[H]
\centering
  \includegraphics[width=\hsize]{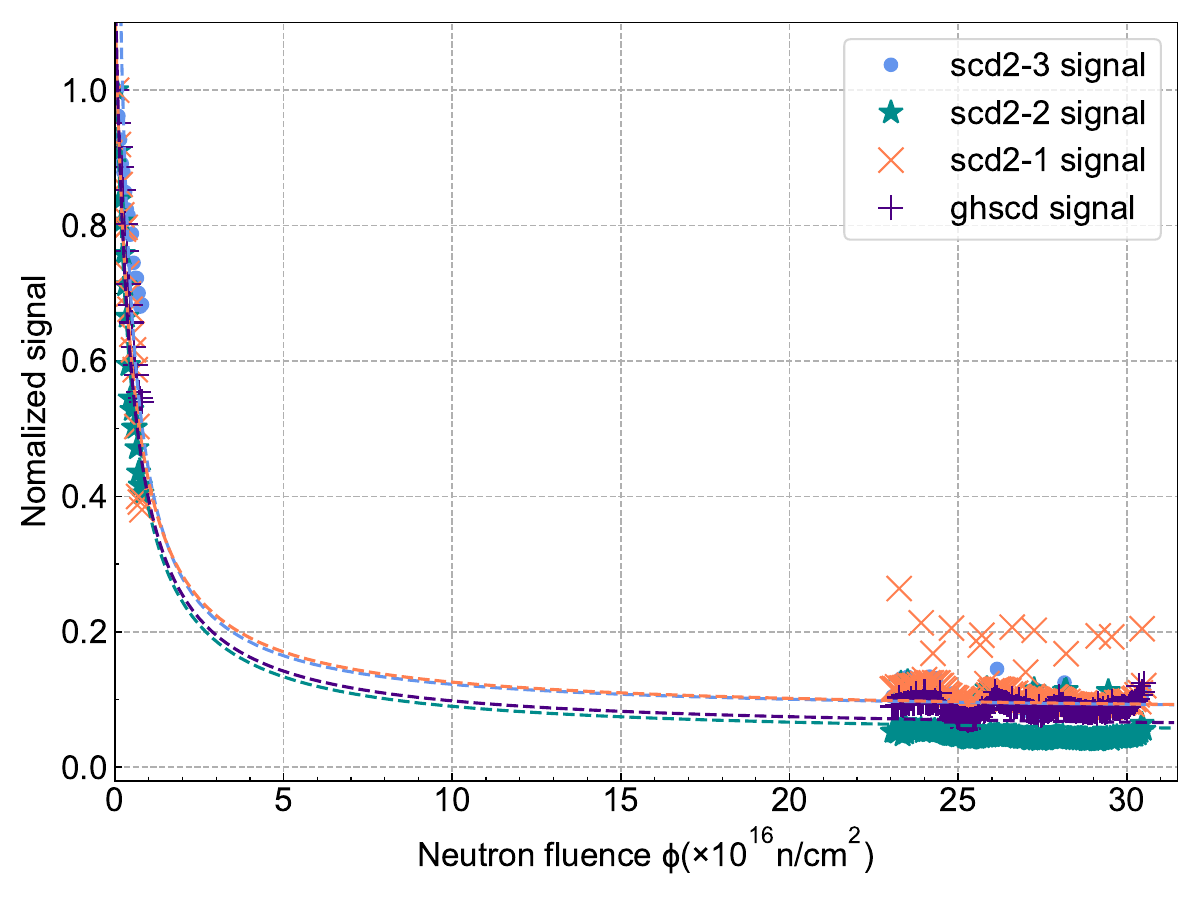}
\caption{Real-time current signal acquisition during neutron irradiation. Beam-off dose contributions have been subtracted. Data acquisition was interrupted during beam recovery, resulting in a missing interval before final stabilization.}
\label{fig3}
\end{figure}

The data acquisition system recorded the continuous DC current response from each of the four single-crystal CVD diamond detector modules under a bias of 260 V, with measurements taken every 30 minutes. As shown in Fig.~\ref{fig2}(b), a sixty hour interruption occurred due to a reactor halted, leading to data interruptions. the relationship between the DC signal and the fast neutron fluence was reconstructed, as presented in Fig.~\ref{fig3}. Following exposure to a fluence of up to \SI{3.3e17}{n/cm^2} fast neutrons, the results indicate that four DUT modules maintained a signal response around 5\% of their initial signal, confirming the modules' robustness against extreme neutron irradiation.
The observed DC signal originates from electron–hole pairs generated by neutron-induced atomic displacements in diamond. Although neutrons are electrically neutral, they induce nuclear collisions governed by a hard-sphere potential, displacing carbon atoms and producing primary knock-on atoms (PKAs) that generate charge carriers. The resulting radiation damage introduces displacement type defects as trapping centers whose density $N_{\mathrm{defects}}$ scales with the accumulated fluence. Assuming that the signal degradation is inversely proportional to the number of traps, the detector response as a function of neutron fluence $\phi$ follows\cite{bib:22}:
\begin{equation}
    S(\phi)=\frac{1}{\frac{1}{S_0}+k\phi}+c
\end{equation}
where $S_0$ is the initial signal, k is the damage constant, and c accounts for baseline offsets, as fitted lines shown in Fig.~\ref{fig3}.
After irradiation, the DC signal remains at 5\% of its initial value with sustained responsiveness, suggesting that sc-CVD diamond is a promising candidate to overcome the lifetime limitations of silicon-based materials in harsh high-radiation environments, particularly in the innermost detector layers of collider experiments.
Signal degradation can be corrected through dedicated calibration procedures. In early simulation studies of the HL-LHC mini-Forward Calorimeters employing CVD diamond sensors, such procedures were devised to account for radiation-induced charge and current losses, thereby ensuring the stability of the energy resolution\cite{bib:31,bib:32}.
\subsection{Radiation induced crystal damage in diamond}
After the irradiation experiment, DUT modules were retrieved from the irradiation facility when their radiation level had sufficiently decreased to an acceptable threshold for human exposure. High dose fast neutron radiation induced damage in sc-CVD diamonds was characterized using photoluminescence (PL) spectroscopy and scanning electron microscope (SEM). For the purpose of investigating crystal damage, sc-CVD diamonds were extracted from the test modules and subjected to a cleaning procedure to remove the metal electrodes. This procedure involved immersing the diamonds in a boiling mixture of hydrochloric acid (HCl) and nitric acid (HNO$_3$), followed by a hydrofluoric (HF) acid treatment to eliminate the Ti-W-Au metal and oxide layer. Additionally, sc-CVD diamonds were sequentially immersed in solutions of acetone, alcohol, and ultra-pure water, respectively, for ultrasonic cleaning to ensure a pristine surface free from any residual organic or inorganic contaminants.
\subsubsection{Luminescence centers associated with radiation-induced defects}
\begin{figure}[H]
\includegraphics[width=\hsize]{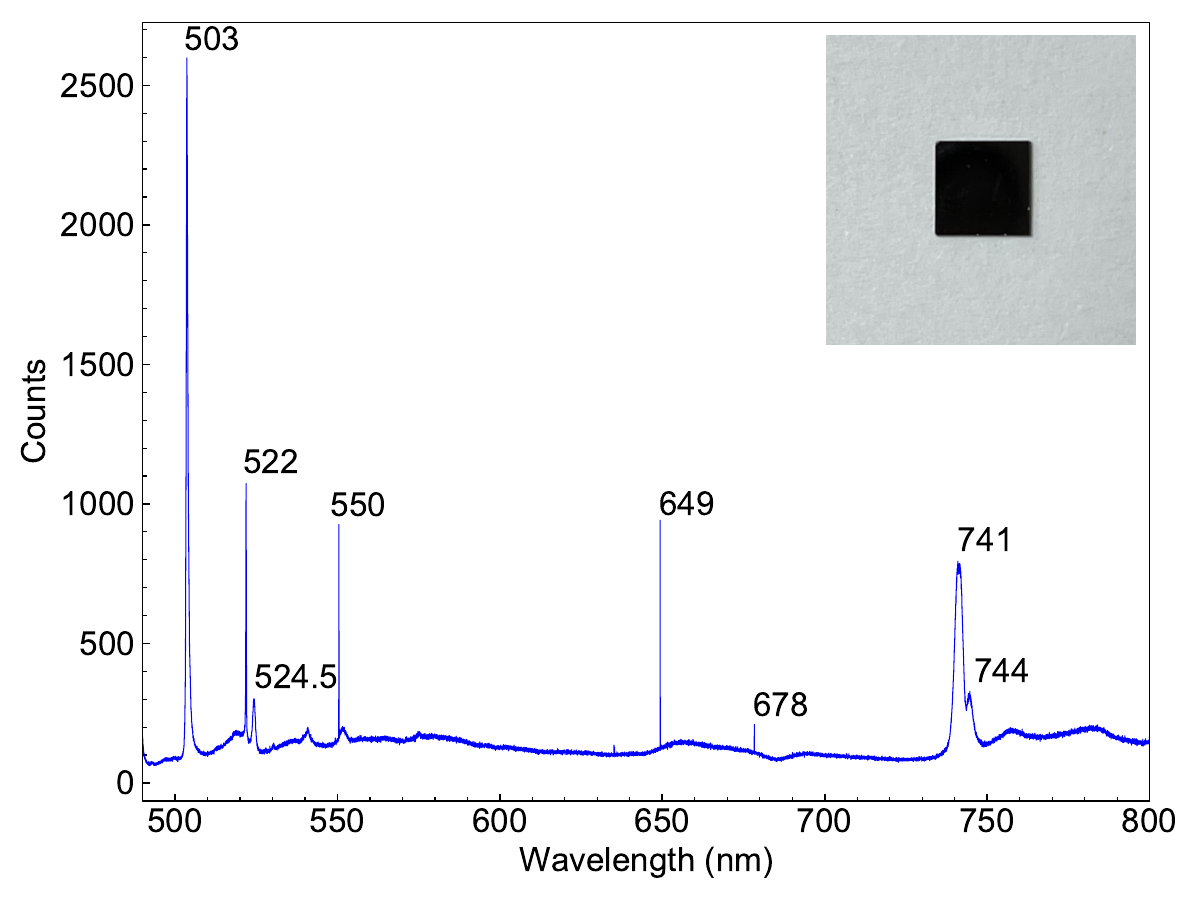}
\caption{Photoluminescence spectroscopy of fast neutron radiated diamond, the upper-right inset shows the optical image after removal of the metal film.}
\label{fig4}
\end{figure}

 The optical image of the sc-CVD diamond after de-metalization and cleaning has been included in the subfigures of Fig. 4. It can be clearly observed that the sc-CVD diamond has turned brown-black, significantly reducing its optical transparency. This phenomenon originates from an increased presence of luminescent centers. Additionally, the incorporation of graphitic regions, nanoscale domains, and interface boundaries into the diamond structure can also contribute to the darkening of diamonds\cite{bib:33,bib:34}. To investigate these radiation induced luminescent centers, PL spectroscopy was performed using a \SI{488}{\nm} blue laser. The temperature was lowered to \SI{77}{\K} in a liquid nitrogen environment, aimed at mitigating the influence of first-order Raman scattering within the diamond.
The PL results of neutron irradiated diamonds are present in the Fig.~\ref{fig4}. Aside from the first-order Raman excitation in \SI{522}{\nm}, additional luminescent centers within the diamond lattice had been induced due to neutron radiation. In Fig.~\ref{fig4}, the \SI{503}{\nm} zero-phonon line (ZPL) can be seen, which originates from two distinct components: the 3H defects\cite{bib:35} and the H3 defects\cite{bib:36,bib:37}, both resulting in a \SI{2.46}{\eV} energy level transition. The 3H defects center consists of single, isolated $\langle 100 \rangle$ self-interstitial defects, a direct consequence of radiation-induced atomic displacement within the diamond lattice. The H3 structure, inclusive of the nitrogen vacancy N-V-N defect, may arise from nitrogen atoms occupying several adjacent vacancies or void defects created by radiation upon the surface of sc-CVD diamond. The presence of the \SI{550}{\nm} center suggests the the occurrence of plastic deformation\cite{bib:38,bib:39} and shear stress\cite{bib:39} in diamond crystal, frequently seen in brown diamonds\cite{bib:40}. The ZPL at \SI{649}{\nm} is assigned to disordered regions in the diamond lattice induced by neutron irradiation\cite{bib:41}. The small ZPL at \SI{678}{\nm} can be refered to nitrogen induced fluorescence peak caused by \SI{1.83}{\eV} photon\cite{bib:42}. The well-known GR1 center in diamond is represented by a ZPL at \SI{741}{\nm}. This center is a common occurrence in diamonds when exposed to radiation and can be attributed to neutral defects of carbon atom vacancies ($\mathrm{V}^0$), which exhibit a tetrahedral ($\mathrm{T_d}$) symmetry. Meanwhile, Jahn-Teller interaction caused by distortion in the GR1 center would result in the degeneracy splitting of the ground state into two sub-levels, $^1E$ and $^1A_1$, with an energy difference of \SI{8}{\meV}. \SI{1.673}{\eV} of energy level transition between lowest ground state 1E and first excited state $^1T_2$ corresponding to the \SI{741}{\nm} ZPL, and the transition involving $^1A_1$ state and $^1T_2$ state leads to the \SI{744}{\nm} side peak, which has a photon energy of \SI{1.665}{\eV}\cite{bib:43}.

\subsubsection{Irradiation-Induced Surface Morphology Characterization}
\begin{figure}[H]
\includegraphics[width=\hsize]{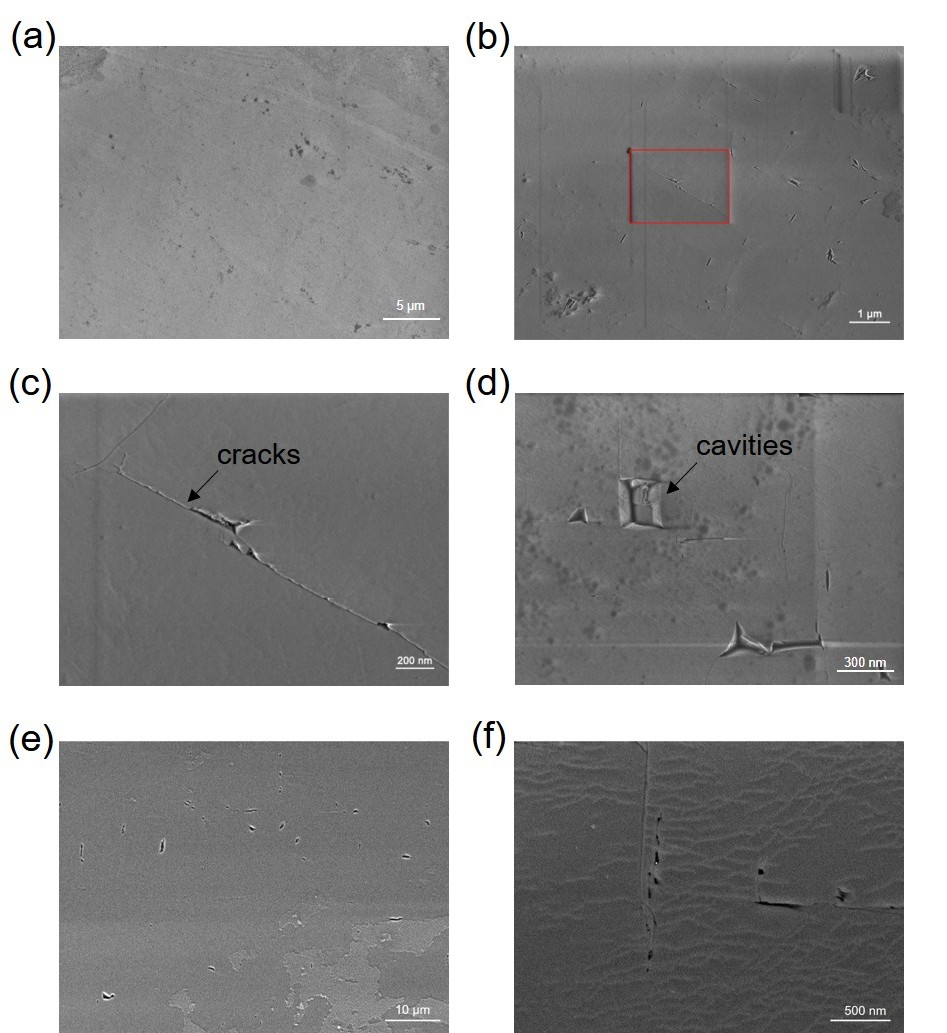}
\caption{Surface morphology of crystals after fast neutron irradiation. (a) Unirradiated diamond surface. (b-d) Surface defect morphology via in-lens secondary electron imaging. (e-f) Surface topography revealed by backscattered electron imaging.}
\label{fig5}
\end{figure}

Scanning electron microscopy (SEM) was employed to investigate the surface morphology of neutron-induced damage in sc-CVD diamond samples. A Zeiss Gemini SEM 500, equipped with both an in-lens detector and a backscattered electron (BSE) detector, was used to acquire high-resolution secondary electron (SE) and backscattered electron images. To enhance sensitivity to surface features and defect contrast, a low accelerating voltage was applied during imaging.
The surface morphologies of the (001) diamond plane after neutron irradiation are shown in Fig.~\ref{fig5}. In comparison with the unirradiated diamond surface in Fig.~\ref{fig5}(a), neutron irradiation has clearly resulted in significant surface damage. Fig.~\ref{fig5}(b)–(d) present secondary electron (SE) images acquired under low accelerating voltage conditions. In particular, Fig.\ref{fig5}(b) reveals intersecting crystal cracks accompanied by cavities distributed along the crack paths. The crystal cracks extend over several micrometres, shown in Fig.\ref{fig5}(c), as the magnified view of the red-boxed region in Fig.\ref{fig5}(b). At the same time, cavities ranging from a few tens to several hundreds of nanometres are evident in the magnified image in Fig.~\ref{fig5}(d).
BSE imaging in Fig.~\ref{fig5}(e) provides further insight into the subsurface damage morphology by revealing the distribution of cavities, as backscattered electrons originate from deeper regions at higher accelerating voltages, according to the Kanaya–Okayama formula\cite{bib:44}.
The magnified BSE image in Fig.\ref{fig5}(f) shows shallow, layered fringe patterns surrounding the cracks, representing the extension of crack-induced damage and indicating the presence of possible stress or graphite layers\cite{bib:45,bib:46} induced by radiation.
Based on SEM analysis, the estimated surface defect densities are approximately \SI{2.9e7}{\per\cm\squared} for cracks and \SI{2.1e7}{\per\cm\squared} for cavities. The cracks, with widths on the order of several nanometres, can be regarded as two-dimensional planar defects analogous to grain boundaries, intersecting the (001) crystal surface. The cavities, in contrast, are thought to be bulk defects emerging at the surface due to the aggregation of vacancies and voids during irradiation, or stress accumulation around point defects leading to localized fracture.

\subsection{Evaluation of the radiation hardness and damage in diamond with damage model}
The irradiation-induced defects in the detector crystal originate from interactions between incident particles and lattice atoms. In this section, we evaluate the impact of crystal damage on the performance of diamond detectors using 100 MeV proton irradiation, and simulate the defect formation process.
\subsubsection{Damage constant of single-crystal diamond under 100 MeV proton with linear damage model}
\begin{figure}[H]
\includegraphics[width=\hsize]{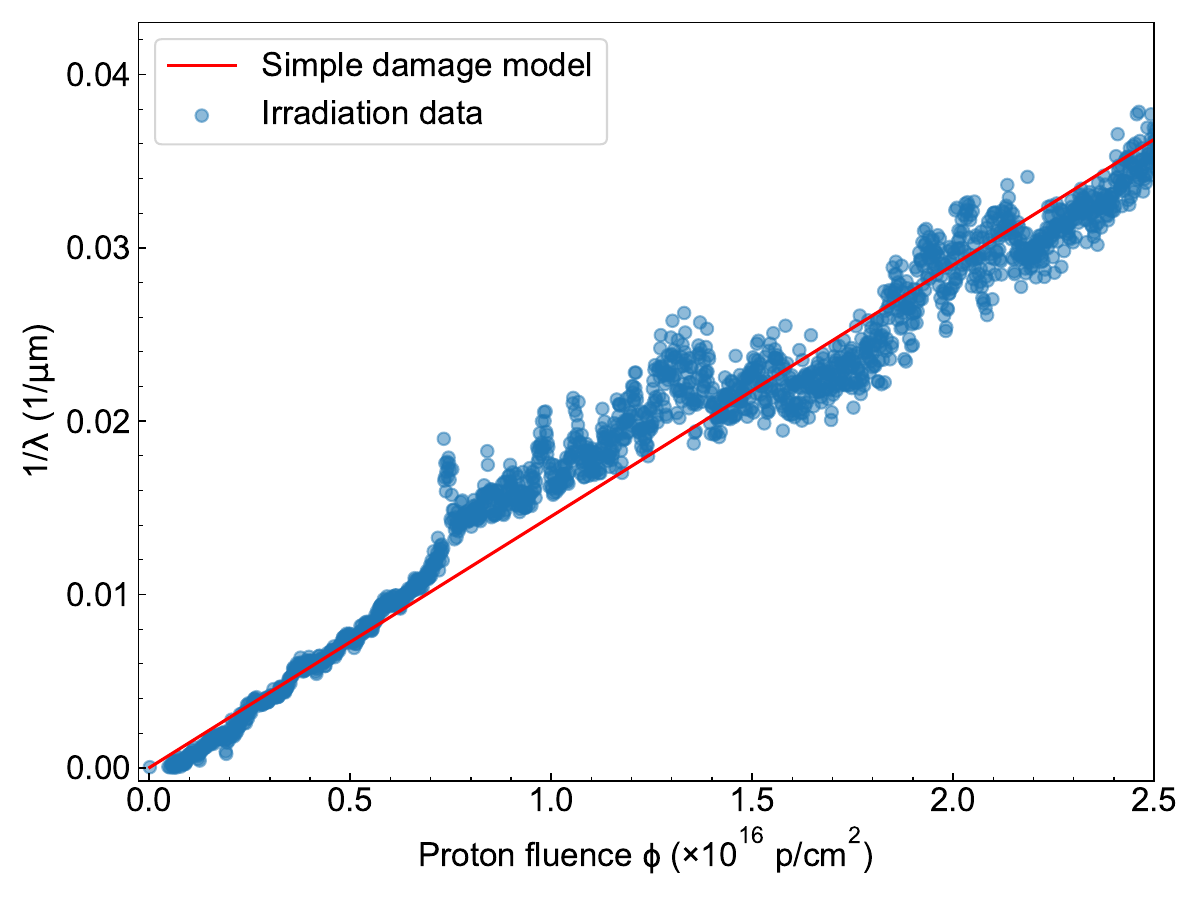}
\caption{100 MeV proton induced charge collection degradation in diamond. The mean free path of carrier decreases linearly with fluence, fitted with red solid line based on linear damage model.}
\label{fig6}
\end{figure}

In diamond detector irradiation experiments, a simplified linear model is often employed to estimate radiation-induced damage\cite{bib:17}. This model assumes that the total number of radiation-induced defects in a crystal scales linearly with the irradiation fluence:
\begin{equation}
 N_{\mathrm{def}} \propto \phi.
\end{equation}
These defects introduce energy levels that trap charge carriers generated by incident particles. Moreover, following a simplified analysis by Kramberger et al.\cite{bib:47}, the carrier lifetime ($\tau$) is inversely proportional to the defect concentration ($\tau \propto 1/N_{\text{def}}$), derived from defect-mediated trapping behavior:
\begin{equation}
\tau_{L}=\frac{1}{\sum_i \sigma_i v_{\mathrm{th}} N_{\mathrm{d,i}}}
\end{equation}
\vspace*{0pt}
Here, $v_{t h}$ denotes the mean drift velocity, $\sigma_i$ represents the interaction (trapping) cross-section for a specific type of charged carrier, and $N_{d,i}$ is the concentration of a given defect/impurity species. Given that the mean free drift distance $1/\lambda$ of carriers scales linearly with their lifetime $\lambda=\tau\times v=\tau\times \mu E$, it follows that the inverse drift distance is proportional to the defect concentration, $1/\lambda \propto N_{def}$, rewrite into linear relationship as \footnote{also know as Messenger-Spratt equation\cite{bib:48,bib:49}: $\Delta\frac{1}{\beta}= \frac{1}{\beta_{\mathrm{irr}}}-\frac{1}{\beta_0}=\frac{1}{\omega_\mathrm{T}}\frac{\Phi_n}{K}$}
\begin{equation}
\frac{1}{\lambda} = k \cdot \phi + \frac{1}{\lambda_0}
\end{equation}
$\lambda_0$ denotes the initial mean free path (MFP) of carriers prior to irradiation, and $k$ is the damage constant, which depends on the type and energy of incident particles, can be expressed as $k(\phi_i,E)$. This parameter quantifies the radiation hardness of a material under specific particle species and monoenergetic flux. As the irradiation fluence increases, radiation-induced damage (e.g., crystallographic defects) accumulates, reducing the MFP and degrading sensor performance (e.g., decreased signal response).
For a given particle type and energy, a smaller k corresponds to slower MFP degradation, indicating superior radiation resistance. Conversely, for a fixed detector, a smaller k under varying particle types or energies implies less damage induced by those irradiation conditions.
In an ideal single-crystal diamond, $1/\lambda_0$ should approach zero due to minimal intrinsic defects, whereas polycrystalline diamonds exhibit a finite $\lambda_0$ owing to grain boundaries and intrinsic defects.
Building on this framework, we utilized results from our previous high-fluence experiment to determine the damage constant of single-crystal diamond under 100 MeV proton irradiation. In previous study\cite{bib:29}, scCVD diamond sensors were subjected to fluence of \SI{2.2e17}{p/cm^2}, with the full set of transient current responses recorded. In planar electrode radiation detectors, the measured charge signal $Q_{meas}$ generated by incident particles is directly related to the mean free path $\lambda_i$ of charge carriers, which can be described using the Hecht\cite{bib:50} model as two representations of charge collection distance (CCD):
\begin{equation}
CCD=\frac{Q_{\mathrm{meas}}}{Q_{\mathrm{gen}}/d}=\sum_{i=e,h}\lambda_i\left[1-\frac{\lambda_i}{d}   (1-e^{-\frac{d}{\lambda_i}})\right].
\end{equation}
where $Q_{gen}$ represents the theoretically generated signal from incident particles, and d is the detector thickness.
Using the signal response current relationship $I=\mathrm{d}Q_{mea}/\mathrm{d}t$, together with the CCD expression in equation (5), we define a calibration factor $\epsilon(\mathrm{d}\phi/\mathrm{d}t,Q_{gen})$ that links the CCD to the measured current $I(\phi)$:
\begin{equation}
 CCD(\lambda(\phi))=\frac{\int I(\phi)\mathrm{d}t}{Q_{gen/d}}=I_{\phi}\cdot \epsilon(\mathrm{d}\phi/\mathrm{d}t,Q_{gen}).
\end{equation}
For high-quality CVD-grown single crystal diamond, the initial CCD can approach the full detector thickness, as demonstrated by the RD42 collaboration\cite{bib:17,bib:51}. In our analysis, we therefore assume an initial CCD close to the detector thickness, i.e., $CCD_0/d\sim 1$. From the initial current response $I_0$, we extract the calibration factor $\epsilon$, which is then applied to irradiated signals $I_{\phi}$ to determine the CCD at each fluence step. The corresponding mean free path $\lambda$ is obtained by solving the CCD expression and subsequently fitted as a function of fluence using the linear degradation model of equation (4). The fitting procedure, illustrated in Fig.~\ref{fig6}, yields the damage constant for single-crystal diamond under irradiation as:
  \begin{equation}
k_{proton}^{100MeV}=1.451\pm0.006(stat)\times 10^{-18} cm^2(p \ \mu m)^{-1}
\end{equation}

To enable direct comparison of radiation damage in diamond across different particle types and energies, a scaling approach\cite{bib:52} was employed. Specifically, the damage constant measured under 100 MeV proton irradiation was normalized to that for 24 GeV protons, yielding a relative damage coefficient $\kappa_\mathrm{i}=k_{\mathrm{p}}(\SI{100}{MeV})/k_{\mathrm{p}}(\SI{24}{GeV})$. A summary of results from this work alongside data from previous studies\cite{bib:17,bib:18} is provided in Table 1. The results show that protons at 100 MeV produce approximately 2.34 times more damage in single-crystal diamond than those at 24 GeV, under equivalent fluence conditions. This scaling enables the estimation of fluence equivalence across irradiation conditions using $\phi_{\mathrm{eq}}=\kappa_{\mathrm{i}} \phi_{\mathrm{i}}$. Such an approach offers a unified framework for quantifying and comparing radiation hardness, allowing the conversion of experimental fluence to equivalent damage levels across a wide range of energies and particle species. 

\begin{table}[H]
\caption{Experimentally fitted radiation damage constants and relative coefficients for 100 MeV protons in diamond.}
\label{tab1}
\begin{tabular*}{8.5cm} {@{\extracolsep{\fill} } lll}
\toprule
Particles & \makecell{Damage\\ constant $k$} & \makecell{Relative \\coefficient $\kappa$}
         \\ \midrule 
        24 GeV proton & 0.62 & 1 \\
        800 MeV proton & 1.04 & 1.67/1.85 \\
        \textcolor{red}{100 MeV proton}\textsuperscript{*} & \textcolor{red}{1.4518} & \textcolor{red}{2.34} \\
        70 MeV proton & 1.55 & 2.48/2.6/2.5 \\
        200 MeV pion & 1.984 & 3.2 \\
\bottomrule
\end{tabular*}
\vspace{2mm} 
\footnotesize \textsuperscript{*}This work.
\end{table}

\subsubsection{Multiscale modelling of radiation damage by combining Monte Carlo and molecular dynamics simulations}
\begin{figure}[H]
\includegraphics[width=\hsize]{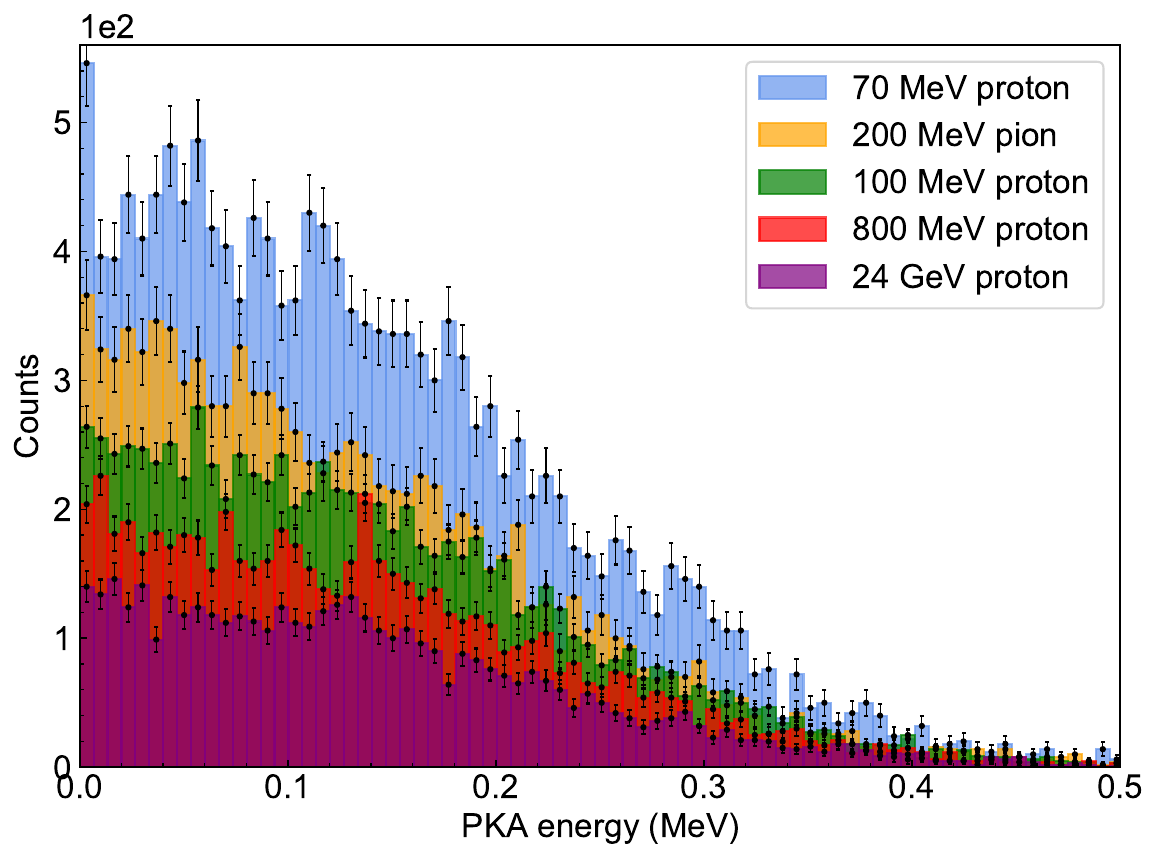}
\caption{Primary knock-on atom (PKA) energy spectra in diamond simulated with Geant4, for incident particles at 24 GeV, 800 MeV, and 100 MeV protons, 70 MeV protons, and 200 MeV pion+.}
\label{fig7}
\end{figure}

To assess the performance degradation of diamond detectors under irradiation, we previously established that crystal damage influences carrier transport properties, ultimately leading to signal attenuation. While radiation-induced lattice damage is often analyzed from the perspective of energy transfer by incoming particles, specifically the non-ionizing energy loss (NIEL) that leads to atomic displacements and phonon excitations\cite{bib:9,bib:53,bib:54}. In this work, we focus on understanding how the diamond crystal lattice itself evolves under irradiation.
In diamond, irradiation damage evolves through a two-stage mechanism. Primary damage occurs when energetic incident particles displace carbon atoms from their lattice sites, producing primary knock-on atoms (PKAs)\cite{bib:55}. These PKAs then initiate secondary collision cascades, generating secondary knock-on atoms (SKAs) and extended structural defects. The distribution of PKA energies plays a pivotal role in determining the final defect morphology and density, thereby controlling the evolution of radiation-induced damage in the crystal.
We employed a hybrid simulation strategy, Monte Carlo (MC) simulations using Geant4\cite{bib:56} were used to compute the PKA energy spectra resulting from interactions between incident particles and the diamond lattice, while molecular dynamics (MD) using LAMMPS\cite{bib:57} was applied to model the subsequent defect cascades induced by PKAs.
As shown in Fig.~\ref{fig7}, PKA energy spectra in diamond were calculated for proton beams with energies of 24 GeV, 800 MeV, 100 MeV, and 70 MeV, as well as for 200 MeV $\pi^+$, revealing how the incident particle type and energy determine the initial damage state in diamond.
Based on Monte Carlo simulations, the probability distribution function $p(E)$ of PKA energy within the interval $E$ to $E+\Delta E$ as
\begin{equation}
p(E)=\rho_N\int_{T_i^\prime}\phi(T_i^\prime)\sigma_\mathrm{d}(T_i^\prime,E)\ \mathrm{d}T_i^\prime\mathrm{d}E
\end{equation}
representing the probability of PKA generation within an infinitesimal energy interval $dE$. Here, $\rho_N$ denotes the atomic number density, $\phi(T_i^\prime)$ represents the incident particle energy spectrum, and $\sigma(T_i^\prime,E)$ is the differential cross-section for producing a PKA with energy $E$ when the incident particle energy is $T_i^\prime$. This process is governed by the interaction cross-sections between incident particles and carbon atoms in diamond, as implemented in GEANT4 simulations.  
According to the law of large numbers,
$\lim_{n\to \infty}\left\{ \left(\frac{f_\mathrm{A}}{n}-p\right)<\varepsilon \right\} =1$
the simulated event frequency converges to the true probability distribution when the sampling size is sufficiently large. Thus, the probability distribution function $p(E)$ can be treated as the actual probability density function $f(E)$. 

Subsequently, the concentration of accumulated displaced atoms resulting from cascade interactions initiated by PKAs is evaluated using the athermal recombination-corrected displacement per atom (arc-DPA) model\cite{bib:58}. This refined framework, which builds upon the traditional Norgett-Torrens-Robinson (NRT)\cite{bib:59} formalism, enables a more accurate estimation of defect production from PKA-induced displacements. The arc-DPA model is formulated as:
\begin{equation}
    N_{\mathrm{d,arc}}(T_{\mathrm{d}})=\left\{\begin{array}{ll}0 & \text { for } T_{\mathrm{d}}<E_{\mathrm{d}} \\ 
    1 & \text { for } E_{\mathrm{d}} \leq T_{\mathrm{d}}<\frac{2 E_{\mathrm{d}}}{0.8} \\ 
    \frac{0.8 T_{\mathrm{d}}}{2 E_{\mathrm{d}}}\cdot\xi_{arc}(T_{\mathrm{d}}) & \text { for } \frac{2 E_{\mathrm{d}}}{0.8} \leq T_{\mathrm{d}}<\infty \end{array}\right.
\end{equation}
$\xi_{\mathrm{arc}}(T_{\mathrm{d}})$ is defined as the efficiency function, expressed as:
\begin{equation}
    \xi_{arc}(T_{\mathrm{d}}) = \frac{1-c_{\mathrm{arc-dpa}}}{(2E_\mathrm{d}/0.8)^{b_{\mathrm{arc-dpa}}}}\cdot T_{\mathrm{d}}^{b_{\mathrm{arc-dpa}}} +c_{\mathrm{arc-dpa}}
\end{equation}
Here, $T_\mathrm{d}$ represents the damage energy, i.e., the kinetic energy of the PKA, while $E_\mathrm{d}$ denotes the average displacement threshold energy of the lattice. The parameters $b_{\mathrm{arc-dpa}}$ and $c_{\mathrm{arc-dpa}}$ are determined through fitting the MD simulations\cite{bib:60}. which describes how the penetration depth of a PKA scales with its energy, and characterizes the saturation value of the defect survival probability within regions of high defect density, particularly relevant at high PKA energies. 

By combining MC-generated spectra (equation 8) within the arc-DPA model (equation 9), we derive the fluence-dependent defect concentration $N(\phi)$ as:
\begin{equation}
    N_{\mathrm{arc}}(\phi)=\int_{E_{\mathrm{pka}}} f(E_{\mathrm{pka}})\cdot N_{\mathrm{arc}}(E_{\mathrm{pka}}) \ \mathrm{d}E_{\mathrm{pka}}
\end{equation}
Once the PKA energy spectra in diamond are obtained for different types and energies of incident particles, it becomes feasible to quantify and compare the resulting radiation damage within the crystal lattice. By correlating the total number of defects with the particle fluence, in a manner analogous to the experimental extraction of the damage constant k, we define a simulation-derived parameter $k_{\mathrm{sim}}$=$N_{\mathrm{arc}}(\phi)/\phi$. This quantity, extracted using a multiscale approach combining MC and MD simulations, characterizes the irradiation-induced degradation of diamond under various radiation conditions.
\begin{figure}[H]
\includegraphics[width=\hsize]{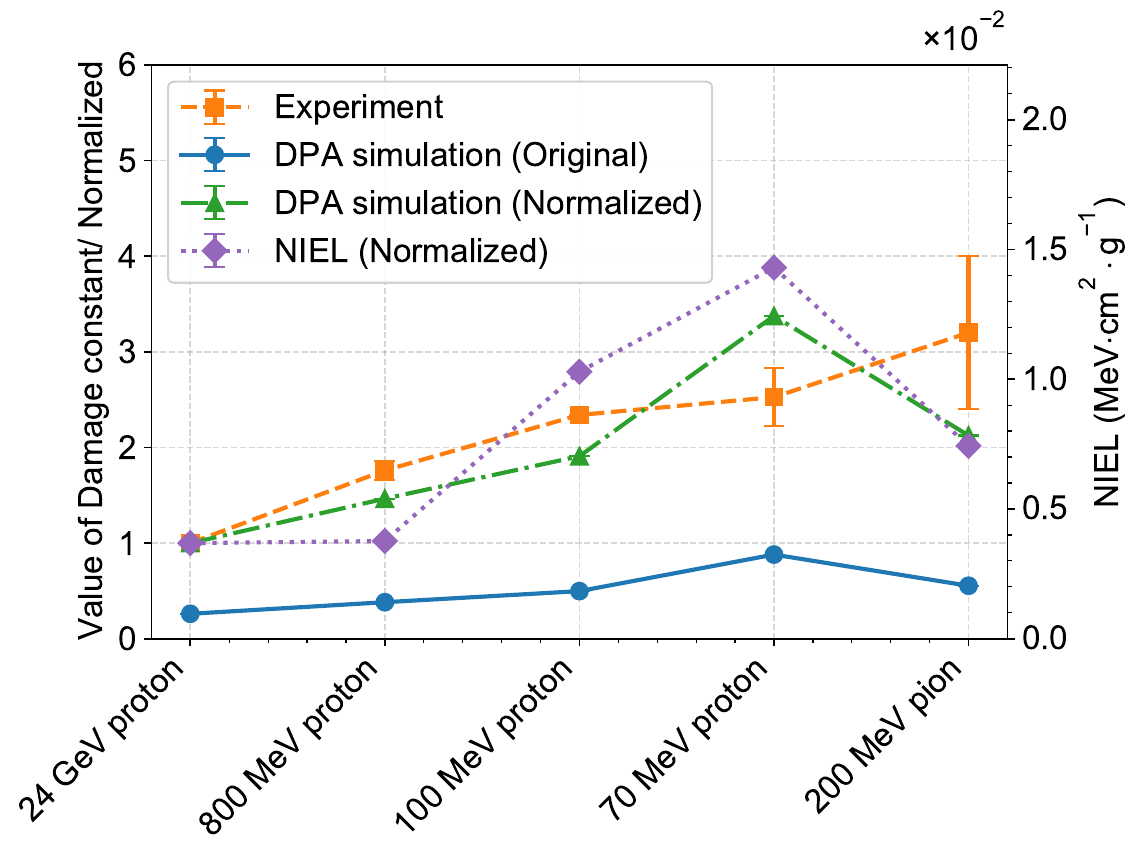}
\caption{Radiation damage constants in diamond for 24 GeV, 800 MeV, 100 MeV, 70 MeV protons, and 200 MeV pion+, comparing experimental measurements (orange), NIEL calculations (purple; from SR-NIEL dataset\cite{bib:61}), and combined Monte Carlo/molecular dynamics predictions (green; arc-DPA model).}
\label{fig8}
\end{figure}

As shown in Fig.~\ref{fig8}, the simulated damage constants $k_{\mathrm{sim}}$ denote in green markers exhibit closer agreement with experimental data than traditional estimates based on the NIEL(purple markers) of the incident particles. This result demonstrates that a crystal-structure-based simulation framework provides a accurate and physically grounded assessment of radiation damage in diamond detectors across different irradiation scenarios.

\subsubsection{Modification of the simple damage model at high radiation dose}
\begin{figure}[H]
\includegraphics[width=\hsize]{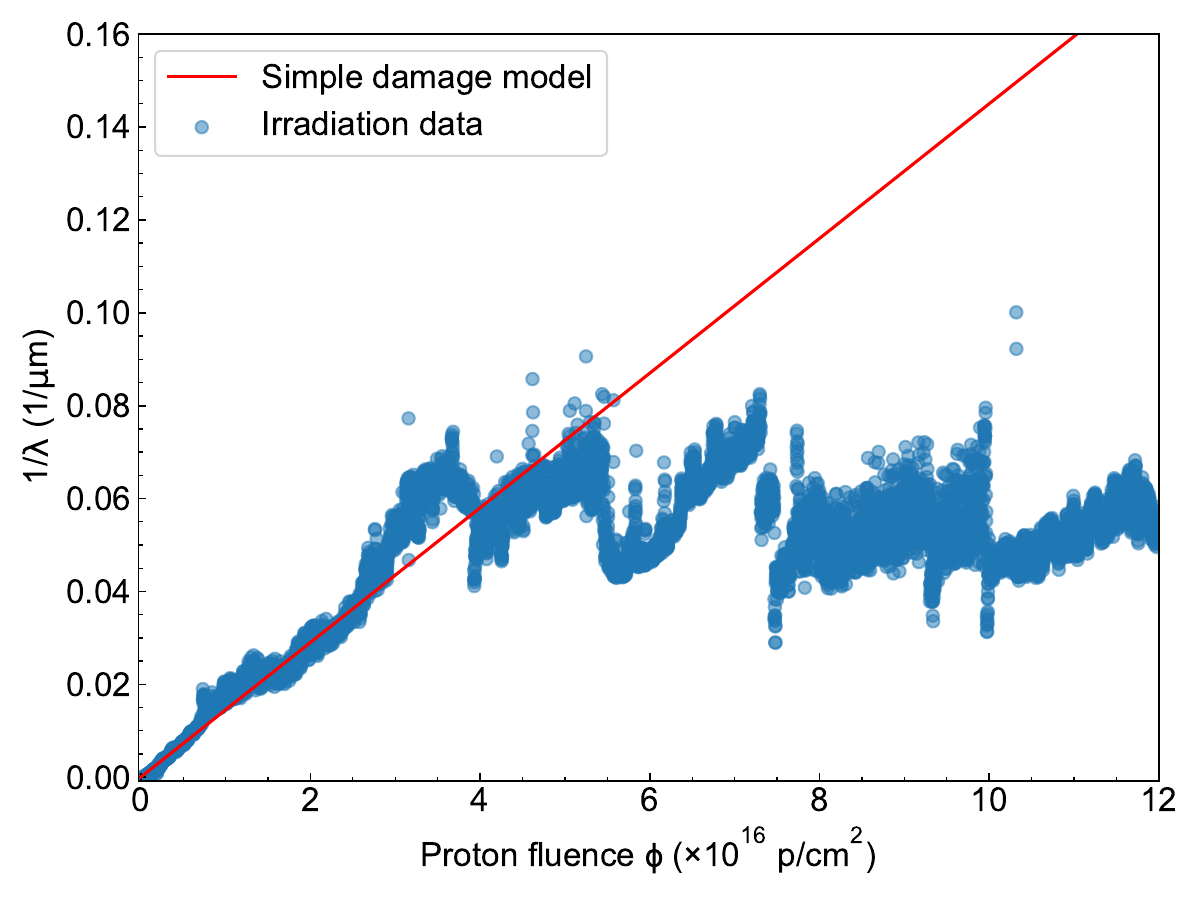}
\caption{Anomaly of linear damage model in diamond under high-fluence 100 MeV proton irradiation. Deviation from linearity initiates at \SIrange{4e16}{5e16}{\per\square\centi\meter}.}
\label{fig9}
\end{figure}

In Section 3.3.1, we applied a linear damage model to describe the degradation of single-crystal diamond under 100 MeV proton irradiation. The model captures the initial trend of radiation-induced performance loss and enables cross-comparison of damage constants. At very high fluences, however, the linear form diverges. In real crystals, atomic density and spatial volume are finite, placing an upper bound on the number of defects that can be generated. At the same time, the continued accumulation of irradiation-induced defects may ultimately drive structural transitions such as amorphization. Under such extreme conditions, the model is expected to break down.
At high doses, irradiation data reveal that for 100 MeV protons, the previously observed linear relationship between carrier mean free path $\lambda$ and particle fluence becomes invalid beyond approximately \SI{4e16}{p/cm^2}. As shown in Fig.~\ref{fig9}, the mean free path begins to saturate, indicating that further damage accumulation no longer leads to proportional degradation in carrier transport. 

In light of the observed deviation behavior, two types of nonlinear damage mechanisms are proposed. The first involves a saturation model of effective defect accumulation. Here, effective defects are defined as those that significantly contribute to the degradation of carrier lifetime, as opposed to the total population of structural disruptions within the crystal lattice.

At high irradiation fluences, where the spatial density of energy deposition events increases significantly within localized regions of the crystal, certain areas may undergo early-stage amorphization or the formation of extended defect clusters, such as the defect clusters and $\mathrm{sp^3-sp^2}$ phase transitions observed in irradiated diamond\cite{bib:60}. Subsequent energetic particles traversing heavily damaged and structurally disordered regions predominantly interact with existing amorphous networks or defect clusters that have already formed, depositing their energy within these disordered structures. In such amorphous domains, the electronic density of states extends into the gap, forming band tails and localized states due to disorder\cite{bib:62}. As a result, additional defects introduced into these regions tend to merge into existing localized states, contributing little further trapping or scattering.
By contrast, when irradiation occurs in regions where the crystal lattice remains relatively ordered, newly generated defect levels act as efficient electrically active traps. These lattice defects are thus identified as the primary contributors of carrier transport degradation.

saturation model of effective defect within a unit volume is proposed, we assume the existence of a saturation defect density $n_{\mathrm{sat}}$, beyond which additional damage has a negligible effect on charge transport. When the local density of effective defects, $n_{\mathrm{eff}}$, remains below $n_{\mathrm{sat}}$, energy deposition by incident particles generates crystal damage that significantly degrades the carrier mean free path. Once $n_{\mathrm{eff}}$ exceeds $n_{\mathrm{sat}}$, further energy deposition in these already disordered or cluster-rich regions is assumed to contribute little additional impact on drift behavior. Across the entire diamond crystal, the local effective defect density $n_{\mathrm{eff}}$ is expressed as a spatially averaged quantity $N_{\mathrm{eff}}$. The evolution of the effective defect density is then described by:
\begin{equation}
N_{\text{eff}} = N_{\mathrm{sat}} \cdot \left(1 - e^{-N(\phi) / N_{\mathrm{sat}}}\right)
\end{equation}
  Using a combined Monte Carlo and molecular dynamics approach, the total defect number can be approximated by the relation $N=N_{\mathrm{arc}}(\phi)=k\phi$, as given by Equation 11. In the low-fluence regime, this expression naturally reduces to the previously linear damage model:
\begin{equation}
\lim\limits_{N\ll N_{\mathrm{sat}}}N_{\text{eff}} \rightarrow N_{\mathrm{sat}} \cdot \left(1 - 1 + \frac{k\phi}{N_{\mathrm{sat}}}\right)=k\phi
\end{equation}
At excessively high fluence conditions, the effective defect concentration approaches a saturation limit:
\begin{equation}
\lim\limits_{N\gg N_{\mathrm{sat}}}N_{\text{eff}} \rightarrow N_{\mathrm{sat}} \cdot \left(1 - 0\right)=N_{\mathrm{sat}}
\end{equation}
By substituting $N_{\text{eff}}$ into Equation 4, we obtain the damage expression under the effective defect model as $1/\lambda = k \cdot N_{\text{eff}} + 1/\lambda_0$. This approach provides a significantly improved description of the degradation behavior at high fluences, as seen by the blue curve in Fig.~\ref{fig13}. From the fitting, we extract the saturation defect density. For 100 MeV protons incident on diamond, this corresponds to approximately 2,036.8 defects within a \SI{100}{nm^3} unit volume as used in Geant4 simulations, which is equivalent to a bulk defect density of \SI{2e18}{/cm^{-3}}. When the number of defects generated by radiation in a local region exceeds this threshold, saturation effects are expected to occur. If we estimate the saturation defect density as $N_{\mathrm{sat}}=N_{\mathrm{arc}}(\phi_0)$, then the onset of saturation effects is expected to emerge at an equivalent fluence $\phi_0$ of approximately \SI{2.5e16}{p/cm^2}.
\begin{figure}[H]
\includegraphics[width=\hsize]{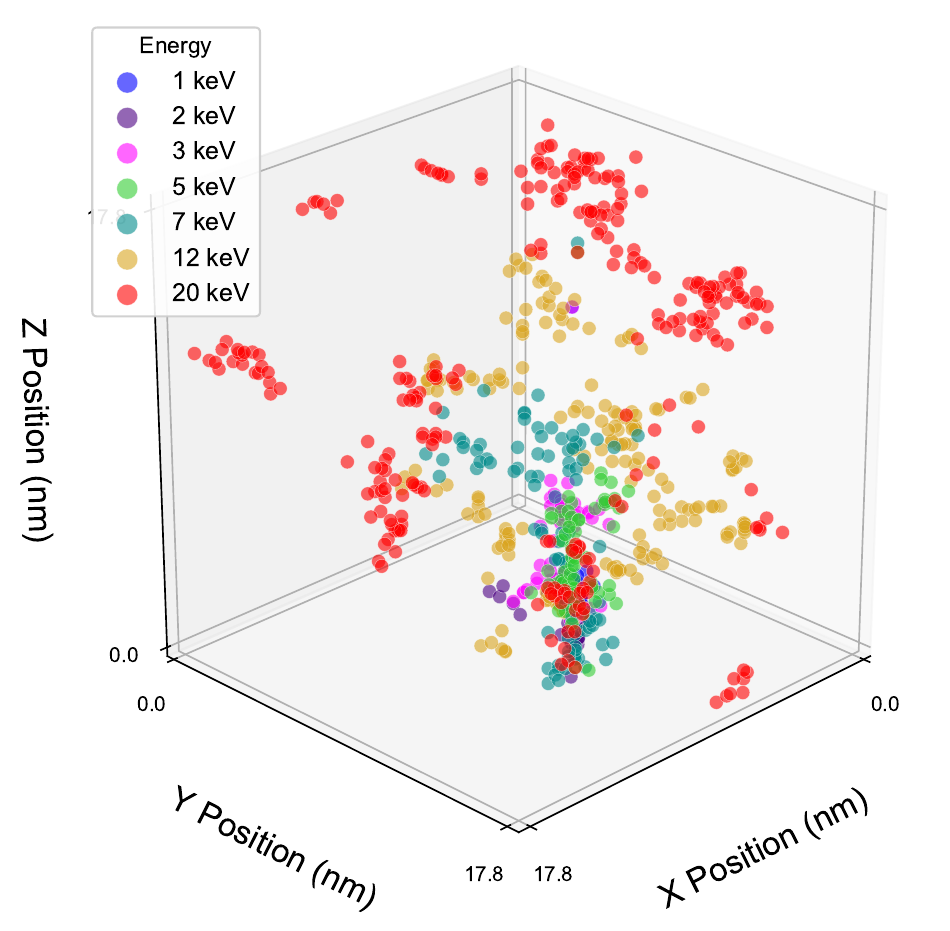}
\caption{Defect distributions in diamond from molecular dynamics simulations of recoil cascades at energies: 1 keV, 2 keV, 3 keV, 5 keV, 7 keV, 12 keV, and 20 keV.  Each marker indicates an individual defect position generated by the cascade process.}
\label{fig10}
\end{figure}
\begin{figure}[H]
\includegraphics[width=\hsize]{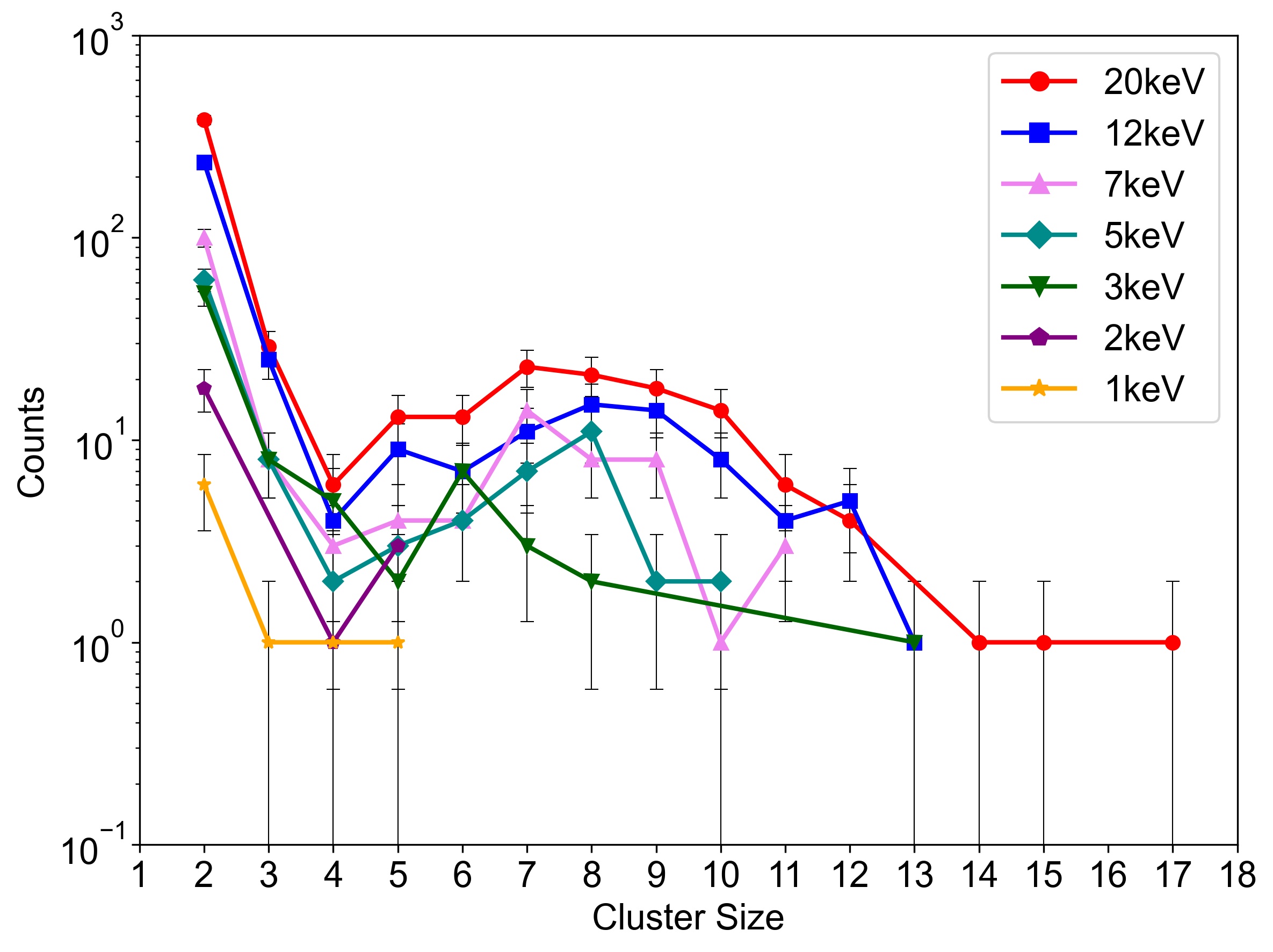}
\caption{Cluster sizes caused by PKAs at different energies in a diamond crystal, measured cut distance of 1.5\AA.}
\label{fig11}
\end{figure}

Inspired by defect interactions and built on the estimation of overlapping irradiation-induced defects, we propose a second nonlinear damage mechanism, as detailed below. Molecular dynamics simulations were employed to estimate the spatial distribution of defects generated by PKAs in diamond. The resulting defect distributions for PKAs with varying recoil energies, obtained using the LAMMPS package, are shown in Fig.~\ref{fig10}. As evident from the visualizations, higher-energy PKAs produce increasingly clustered defects. To quantify this behavior, we define a cutoff distance of \SI{1.5}{\angstrom} between carbon atoms and use it to statistically evaluate the size and number of defect clusters. The results are presented in Fig.~\ref{fig11}. These simulations reveal that a single PKA not only forms localized defect clusters but also establishes a characteristic spatial range over which its damage extends.

\begin{figure}[H]
\includegraphics[width=\hsize]{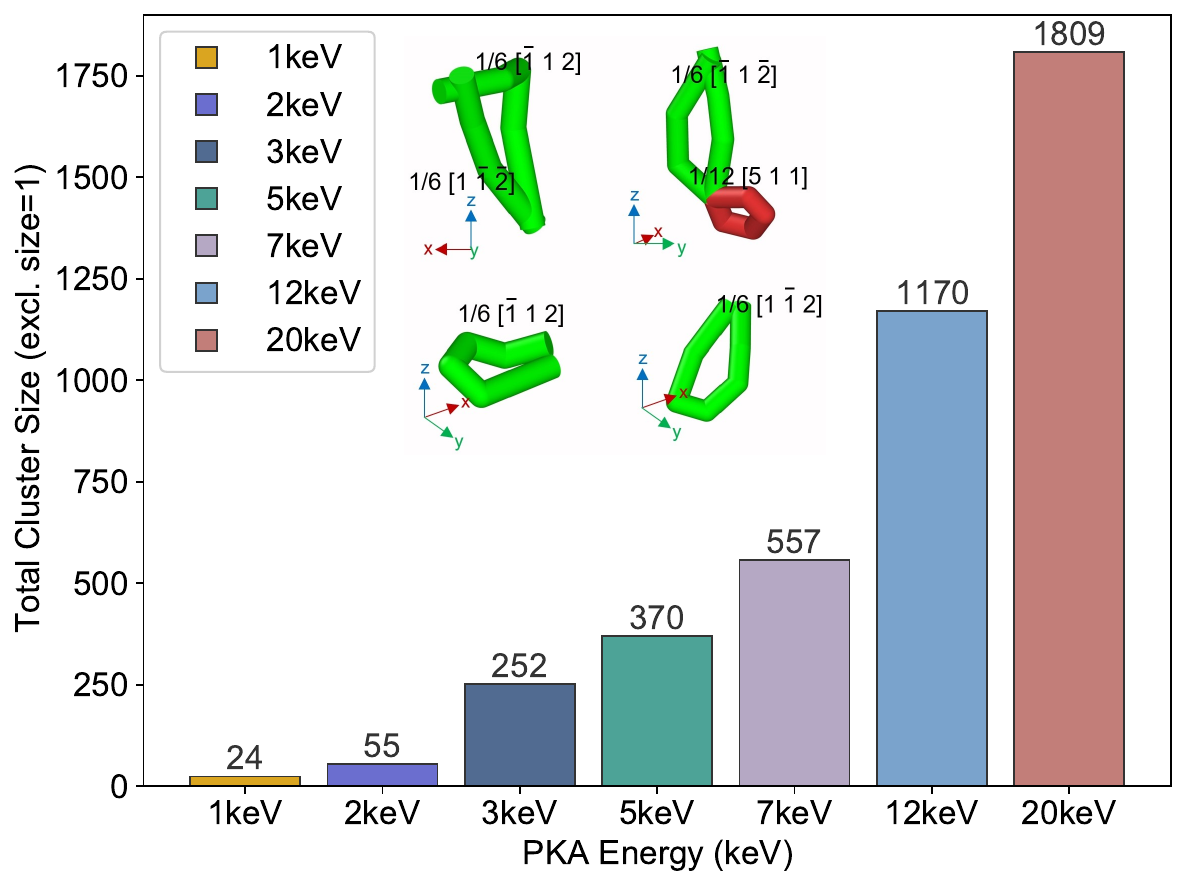}
\caption{ Cumulative defect clusters and dislocations induced by PKAs at different energies in diamond crystal (excluding counts of 1 Wigner-Seitz unit cell size).}
\label{fig12}
\end{figure}

\begin{figure}[H]
\includegraphics[width=\hsize]{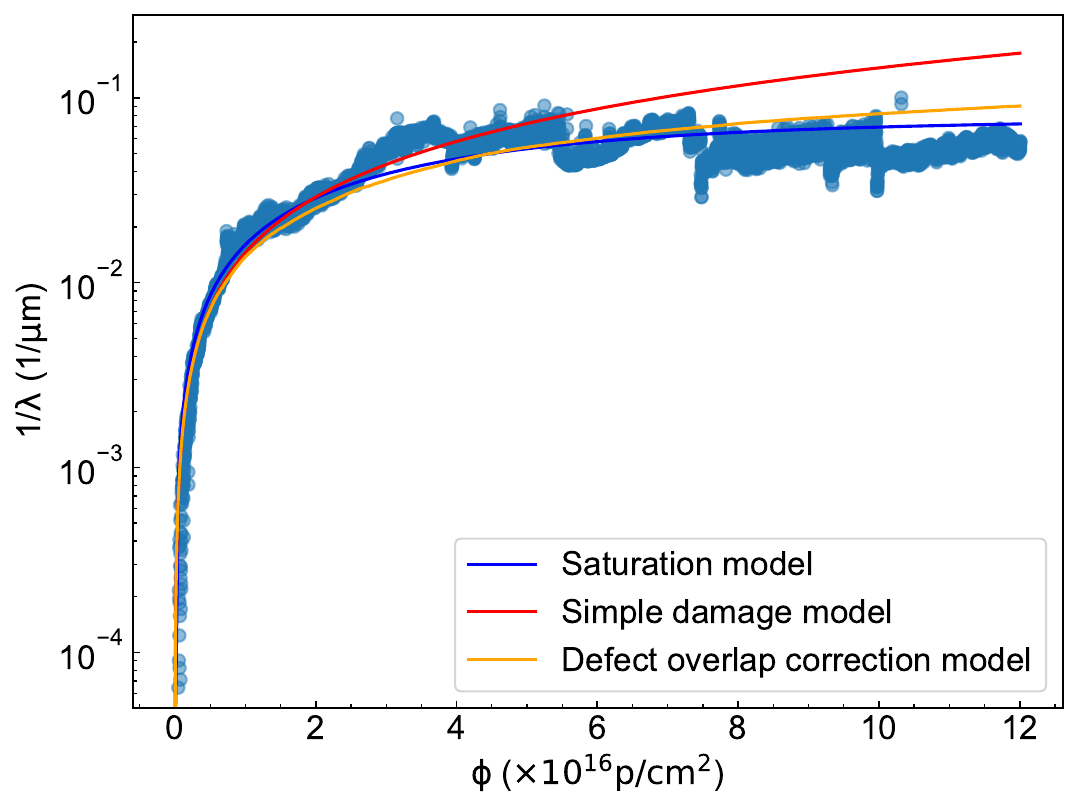}
\caption{Modification of the linear damage model at high fluences. The degradation behavior is better described by incorporating effective defect saturation (blue curve) and defect-overlap effects considering interactions between defects (orange curve).}
\label{fig13}
\end{figure}

As shown in Fig.~\ref{fig12}, the cumulative number of clusters produced by a PKA increases with recoil energy, and dislocation defects are also observed in the simulations. These findings indicate that each PKA generates a distinct damage volume comprising energy-dependent cluster regions and extended defects. At high fluences, the probability that a new PKA’s damage volume overlaps with pre-existing damage regions rises. We hypothesize that within these overlapping regions, interactions between clusters dominate over the formation of isolated point defects. Furthermore, we assume that such overlapping volumes contribute minimally to additional carrier degradation. Consequently, only non-overlapping PKA damage regions are considered effective in reducing carrier transport. We define the effective damage volume as $V_{\mathrm{eff}}(\phi)=V_\mathrm{tot}(\phi)-V_{\mathrm{overlap}}(\phi)$, and introduce a fluence-dependent damage reduction factor: 
\begin{equation}
k_{\mathrm{overlap}}(\phi) = \frac{V_{\mathrm{tot}}-V_{\mathrm{overlap}}}{V_{\mathrm{tot}}}
\end{equation}

In the Monte Carlo simulations, the spatial positions of individual PKAs were recorded. The total damage volume, $V_{\mathrm{tot}}$, was obtained by summing the effective defect volumes produced by each PKA over the entire irradiation period. Simultaneously, the overlapping volume, $V_{\mathrm{overlap}}$, was determined by calculating the intersection between newly generated PKA defect volumes and those already existing. For each PKA, the associated damage volume was derived by fitting molecular dynamics simulation results (Fig.~\ref{fig10}) for various recoil energies. Finally, by replacing the constant damage parameter in Eq. (4) with the fluence-dependent damage reduction factor at each fluence, we obtained the carrier mean free path as a function of fluence according to the second nonlinear damage mechanism, as shown by the orange curve in Fig.~\ref{fig13}. This modified model, incorporating defect overlap, reproduces the damage behavior observed at high fluence.

\begin{figure}[H]
\centering
\subfigure[]{
  \includegraphics[width=0.149\textwidth]{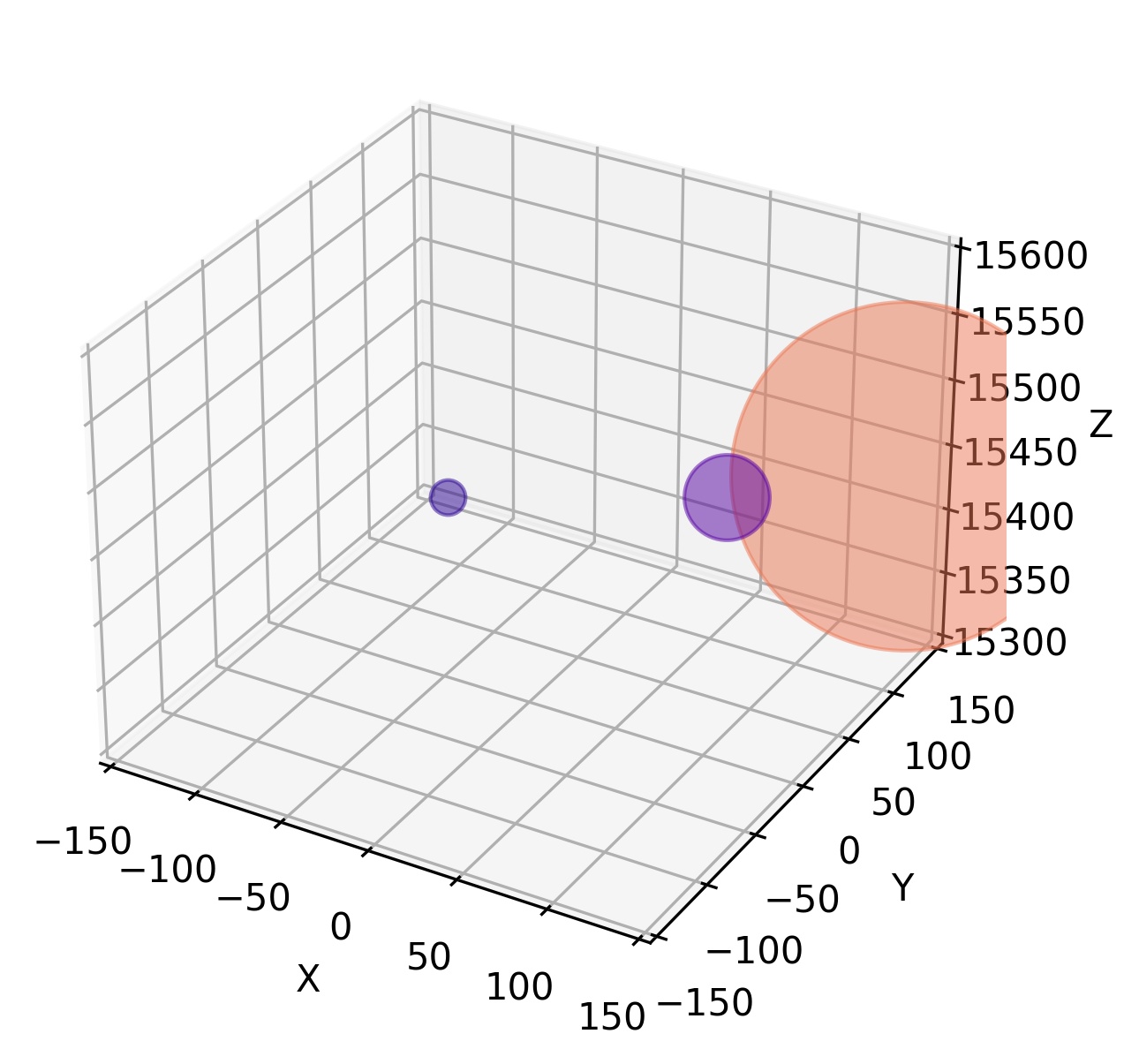}
}
\subfigure[]{
  \includegraphics[width=0.149\textwidth]{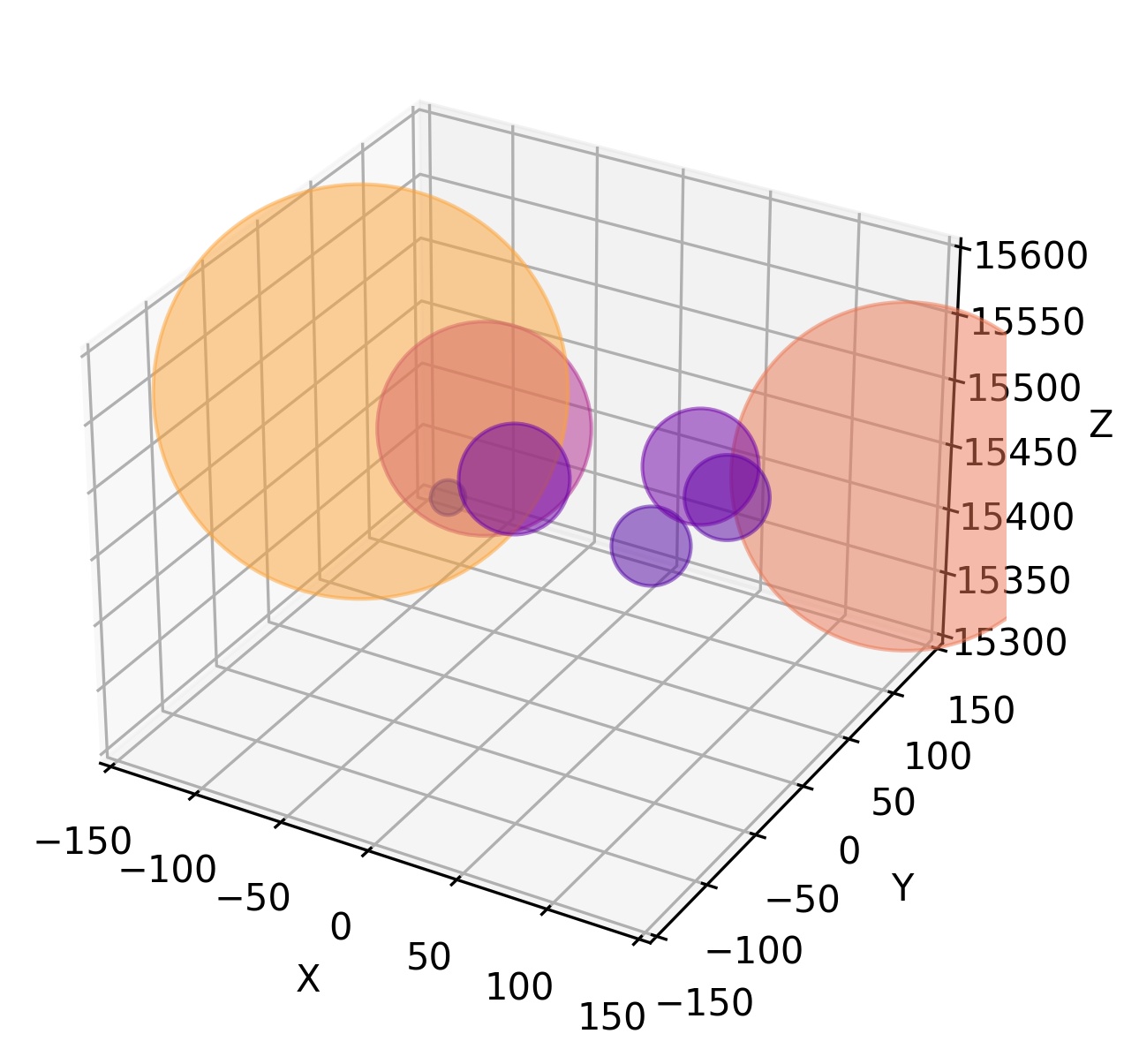}
}
\subfigure[]{
  \includegraphics[width=0.149\textwidth]{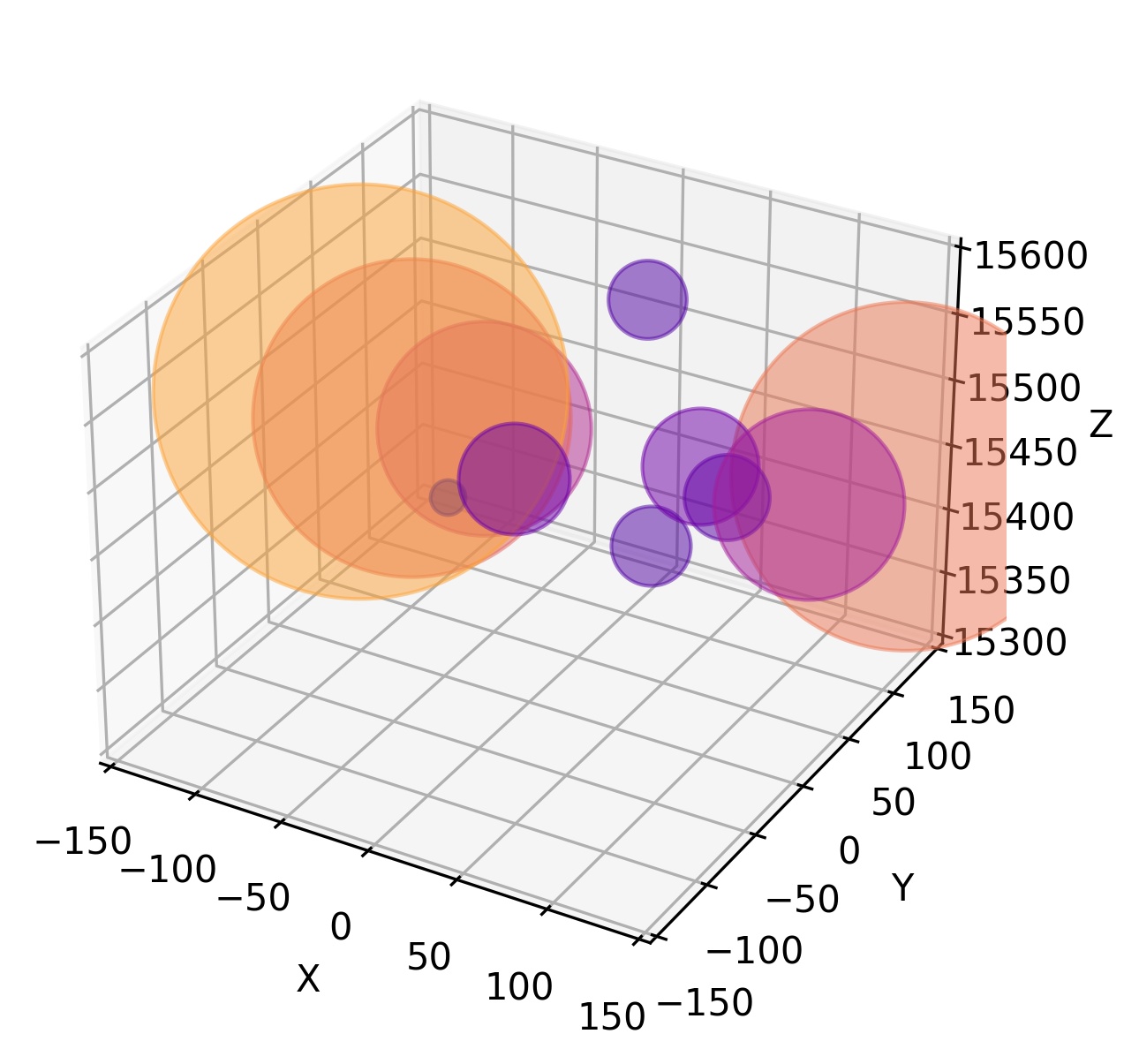}
}
\subfigure[]{
  \includegraphics[width=0.149\textwidth]{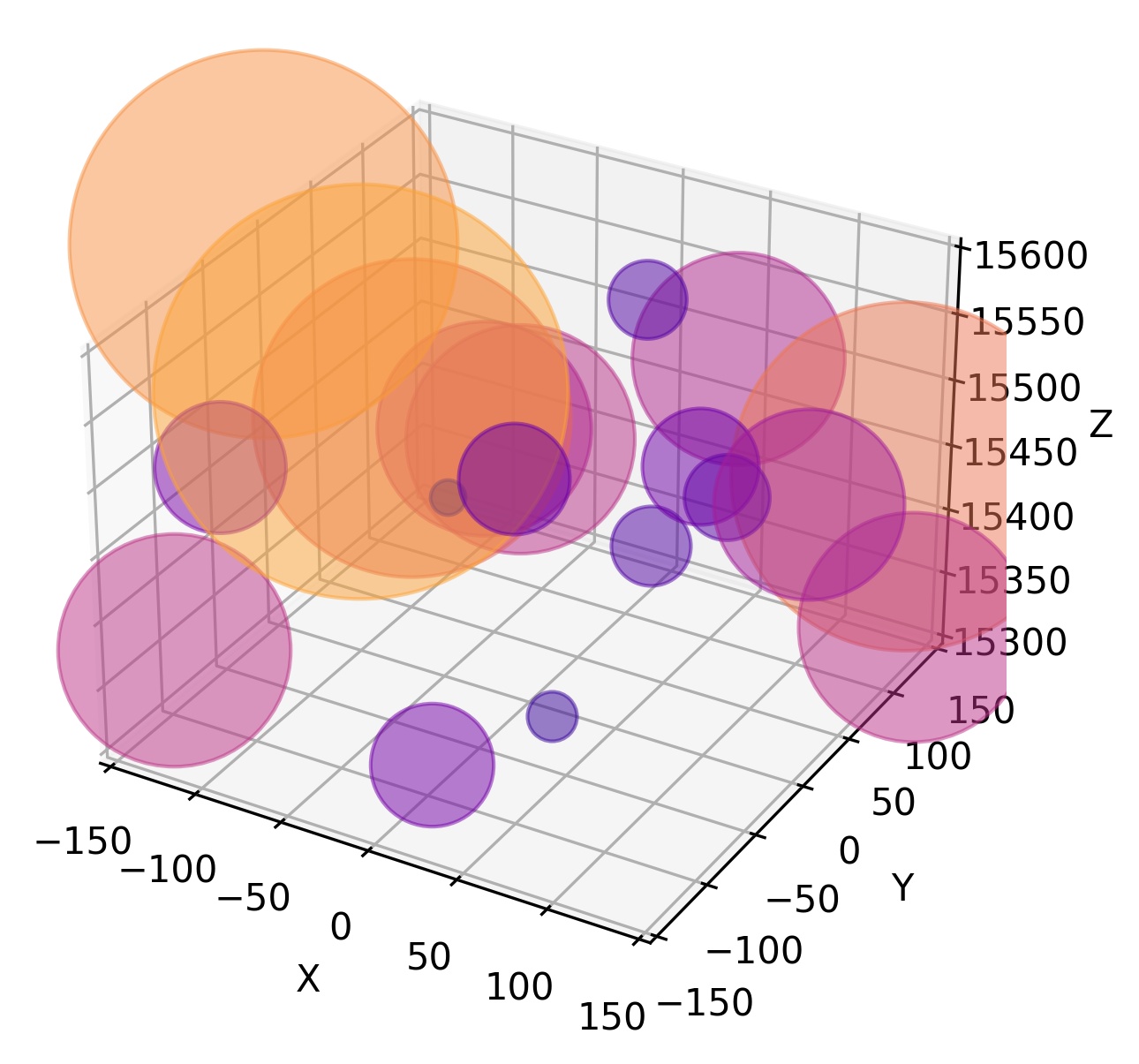}
}
\subfigure[]{
  \includegraphics[width=0.149\textwidth]{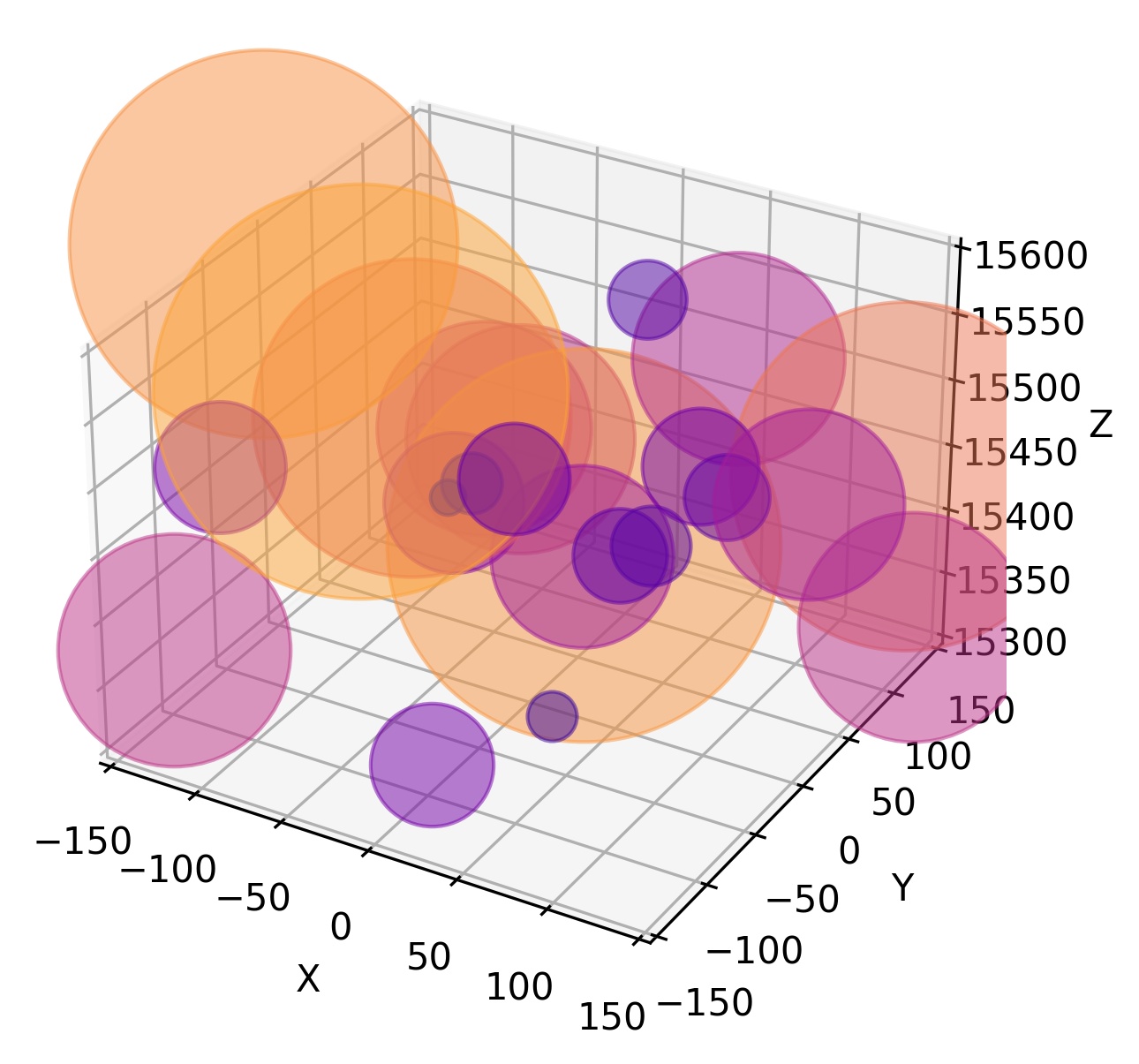}
}
\subfigure[]{
  \includegraphics[width=0.149\textwidth]{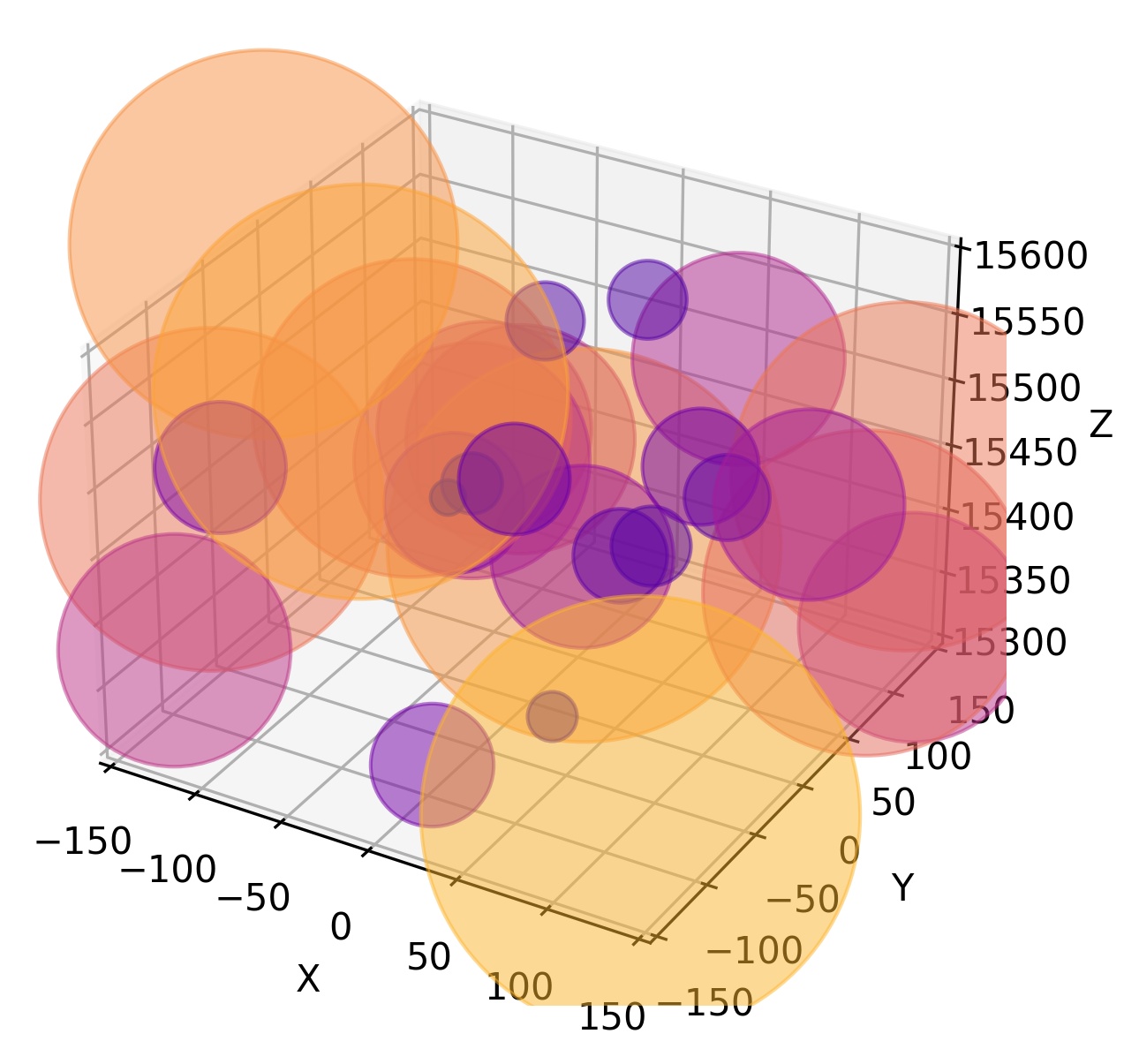}
}
\subfigure[]{
  \includegraphics[width=0.149\textwidth]{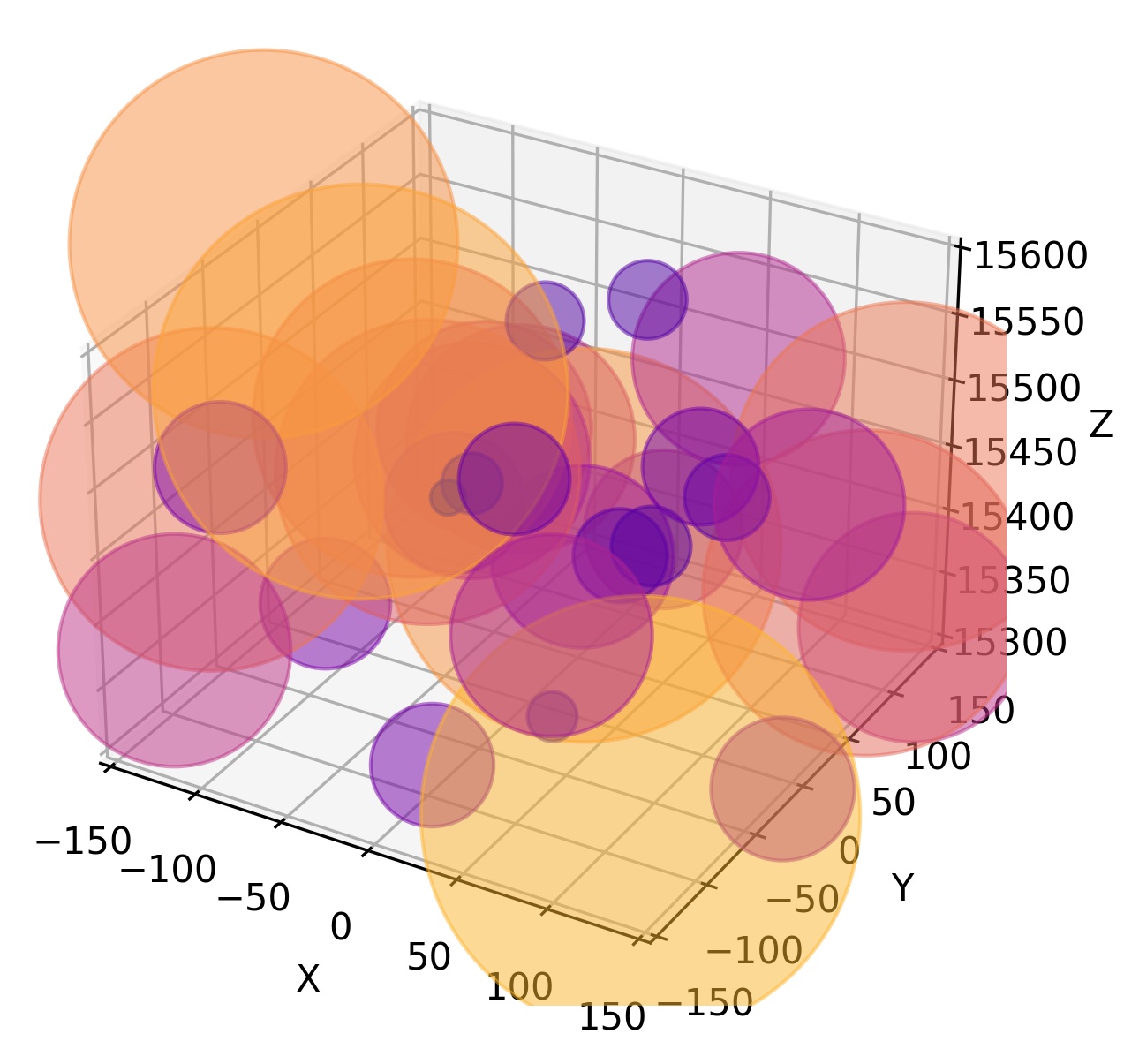}
}
\subfigure[]{
  \includegraphics[width=0.149\textwidth]{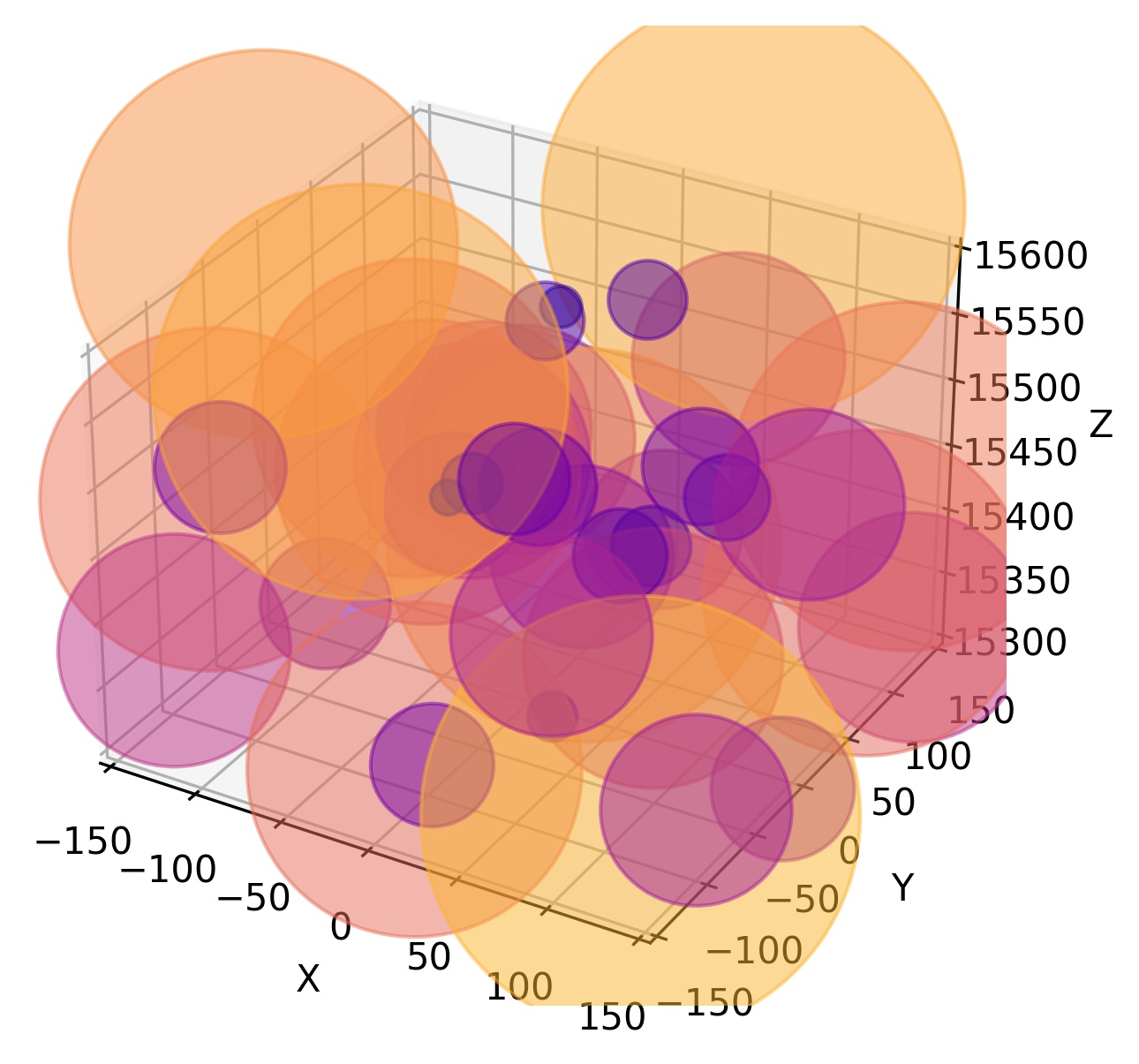}
}
\subfigure[]{
  \includegraphics[width=0.149\textwidth]{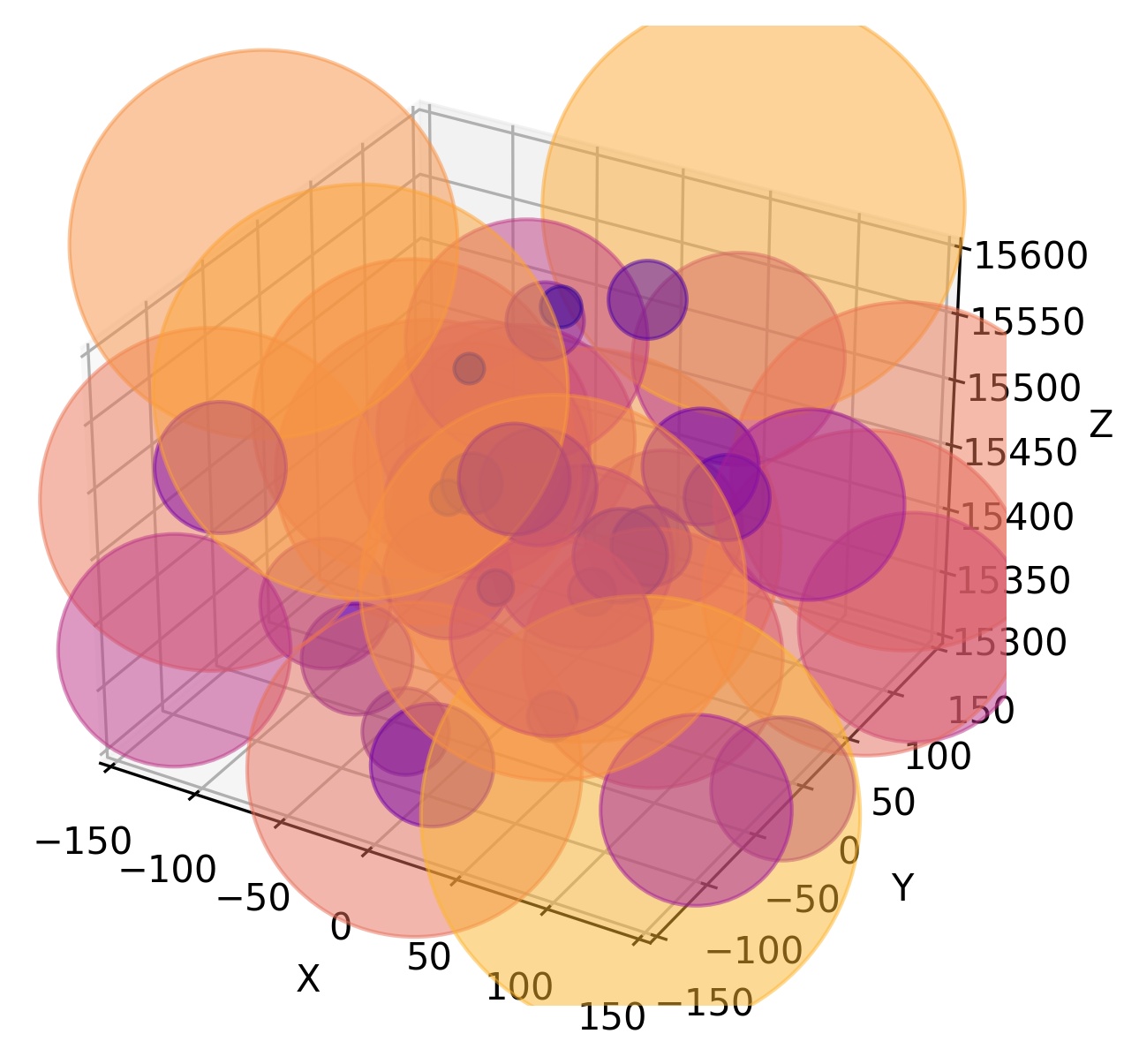}
}
\subfigure[]{
  \includegraphics[width=0.149\textwidth]{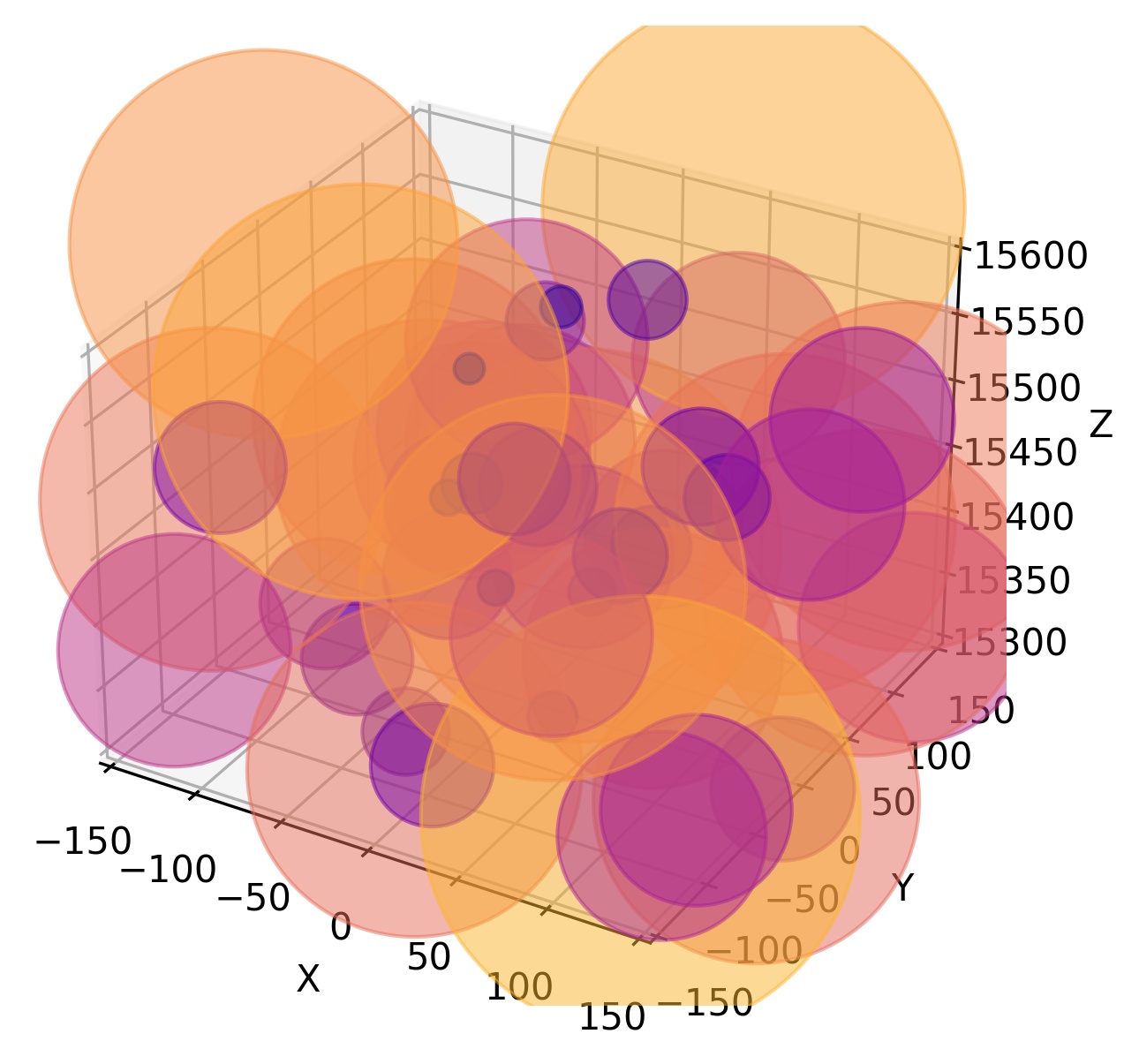}
}
\caption{(a)–(j) show the spatial distribution of PKA induced defect coverage in diamond under 100 MeV proton irradiation at fluences of $\phi=1.2 \times 10^{16}$, 
$\phi=2.4 \times 10^{16}$, 
$\phi=3.6 \times 10^{16}$, 
$\phi=4.8 \times 10^{16}$, 
$\phi=6.0 \times 10^{16}$, 
$\phi=7.2 \times 10^{16}$, 
$\phi=8.4 \times 10^{16}$, 
$\phi=9.6 \times 10^{16}$, 
$\phi=1.08 \times 10^{17}$, 
$\phi=1.2 \times 10^{17}$, based on effective defect volumes derived from MD simulations.}
\label{fig14}
\end{figure}

To further illustrate this effect, we consider a diamond crystal with a volume of 
\SI{300}{nm^3}. At different proton fluences, denoted as $\phi=1.2 \times 10^{16}$, 
$\phi=2.4 \times 10^{16}$, 
$\phi=3.6 \times 10^{16}$, 
$\phi=4.8 \times 10^{16}$, 
$\phi=6.0 \times 10^{16}$, 
$\phi=7.2 \times 10^{16}$, 
$\phi=8.4 \times 10^{16}$, 
$\phi=9.6 \times 10^{16}$, 
$\phi=1.08 \times 10^{17}$, 
$\phi=1.2 \times 10^{17}$, the number of PKAs and their corresponding spatial damage ranges are visualized in Fig.~\ref{fig14}. In the figure, the circular mappings represent the extent of defect-affected regions, while different colors correspond to the defect volumes generated by different PKAs. Lighter colors indicate larger affected volumes. It is evident that as the fluence exceeds around \SIrange{3e16}{4e16}{p/cm^2}, the overlapping regions between PKA-induced damage volumes increase significantly, leaving less undisturbed crystal volume for the formation of isolated point defects. From the perspective that each PKA generates multiple defect clusters, the increasing dominance of defect–defect interactions becomes apparent at high fluences.

Through the effective defect saturation model and the defect interaction model, this study establishes a connection between irradiation-induced defect evolution and carrier transport degradation in diamond, providing a phenomenological framework that captures the nonlinear behavior of diamond detectors under high radiation fluence.
The first model assumes a saturation of electrically active defects capable of trapping carriers, while the second considers the spatial overlap and interaction among newly generated and pre-existing defects. Together, they account for the observed deviation from linearity in detector response at high doses.
Both frameworks converge on a consistent interpretation that under intense irradiation the diamond lattice undergoes a structural transformation from a regime dominated by isolated point defects to one governed by defect clusters and locally amorphous configurations. This structural transition in radiation damage modifies the distribution of density of states within the band structure, subsequently changing the effective carrier trapping cross section and driving the detector response from linear to nonlinear behavior in CCD and other electrical characteristics.
Monte Carlo and molecular dynamics simulations indicate that this transition begins to emerge when the defect density approaches \SI{e18}{/cm^{-3}}, corresponding to an equivalent neutron fluence of \SI{e16}{n/cm^3}, where linear scaling with fluence no longer holds. In this regime, comprehensive understanding of these nonlinear processes will require integrated first principles calculations and advanced experimental probes capable of resolving the atomic-scale mechanisms underlying defect clustering and amorphization in diamond.

\section{Conclusion}
This study demonstrates that single-crystal CVD diamond sensors retain functional signal response under exceptionally high radiation fluence, maintaining approximately 5\% of their initial output after fast neutron irradiation up to \SI{3.3e17}{n/cm^2} among the highest levels tested to date. These results confirm the feasibility of applying such sensors in extreme radiation environments.

Spectroscopic and electron microscopy analyses revealed that both bulk and surface defects induced by irradiation, such as self-interstitials, vacancies, and nanoscale surface cracks play a central role in the observed degradation of detector performance. Following the linear carrier-drift degradation framework, we experimentally extracted the quantitative damage constant for 100 MeV protons based on low fluences experimental data as $k^{\SI{100}{MeV}}_{\mathrm{proton}}$ = \SI{1.452\pm0.006 e-18}{cm^2/(p.\um)}, providing essential reference data for radiation damage assessment in diamond under medium-energy proton irradiation conditions.

To explore the underlying mechanisms of damage saturation, we performed multiscale simulations that couple Monte Carlo particle transport with molecular dynamics modeling. This combined approach yields damage estimates that align more closely with experimental observations than conventional NIEL predictions, offering a new perspective for studying radiation damage in diamond detectors. Furthermore, the framework provides a phenomenological means to investigate the observed saturation in carrier transport at high doses and to refine nonlinear degradation models. At high fluences, interactions between defects may become the dominant mechanism of lattice modification, gradually replacing isolated point defect formation. This transition leads to the emergence of a local effective saturation defect density, beyond which additional damage has a diminishing effect on the carrier drift length.

In summary, these findings establish a fundamental understanding of radiation-induced damage in diamond at high fluence, and offer practical guidance for the design and deployment of diamond-based detectors in future high-radiation particle physics experiments and advanced nuclear technologies.

\section*{Acknowledgements}
The authors would like to acknowledge M. Bulavin, A. Cheplakov, L. Kurchaninov, V. Kukhtin, J. Ye, C. Li, H. Ding, J. Du, Z. Zou and F. Miao for their testing facilities and helpful discussions. They would also like to acknowledge the ATLAS-LAr Collaboration for their strong supports and beneficial comments. This work was supported by the “International Science \& Technology Cooperation Program of China”(Contract No. 2015DFG02100), The Ministry of Science and Technology of the People’s Republic of China.

\section*{Author contributions}
All authors contributed to the work. Jialiang Zhang, Shuo Li, Guojun Yu, and Zifeng Xu performed characterization experiments, data analysis, simulations, and manuscript writing. Yilun Wang and Shuxian Liu contributed to model analysis. Lifu Hei and Fanxiu Lv provided materials. Ming Qi contributed to experimental design, planning, conducted irradiation experiments, and provided funding support. Ming Qi and Lei Zhang were responsible for manuscript revision and supervision. The first draft of the manuscript was written by Jialiang Zhang, and all authors commented on previous versions. All authors have read and approved the final manuscript.

\section*{Data availability}
The data that support the findings of this study are available from the corresponding author upon reasonable request.

\section*{Declaration of interests}
The authors declare that they have no known competing interests or personal relationships that could have appeared to influence the work reported in this paper.

\end{document}